\documentclass[epj]{svjour}
\usepackage{graphics}
\begin{document}
\title{Symmetry energy, unstable nuclei, and neutron star crusts}
\author{Kei Iida\inst{1,2} \and Kazuhiro Oyamatsu\inst{2,3}% etc
}                     % Do not remove
\institute{Department of Natural Science, Kochi University,
Akebono-cho, Kochi 780-8520, Japan \and
RIKEN Nishina Center, Wako-shi, Saitama 351-0198, Japan
\and Department of Human Informatics, Aichi Shukutoku
University, 9 Katahira, Nagakute, Aichi 480-1197, Japan}
\date{Received: date / Revised version: date}
\abstract{
Phenomenological approach to inhomogeneous nuclear matter is useful to 
describe fundamental properties of atomic nuclei and neutron star crusts
in terms of the equation of state of uniform nuclear matter.  We review
a series of researches that we have developed by following this approach.
We start with more than 200 equations of state that are consistent with
empirical masses and charge radii of stable nuclei and then apply them
to describe matter radii and masses of unstable nuclei, proton elastic 
scattering and total reaction cross sections off unstable nuclei, and 
nuclei in neutron star crusts including nuclear pasta.  We finally discuss
the possibility of constraining the density dependence of the symmetry 
energy from experiments on unstable nuclei and even observations of
quasi-periodic oscillations in giant flares of soft gamma-ray repeaters.
 % end of PACS codes
} %end of abstract
\maketitle
\section{Introduction}
\label{intro}
%Your text comes here. Separate text sections with
%\section{Section title}
%\label{sec:1}
%and \cite{RefJ}
%\subsection{Subsection title}
%\label{sec:2}
%as required. Don't forget to give each section
%and subsection a unique label (see Sect.~\ref{s%ec:1}).
%

Determining the equation of state (EOS) of uniform nuclear matter is an old 
and fundamental issue in nuclear physics, but it is rather hard to solve 
\cite{HP}.  
Thus, it is still important to keep studying the EOS of nuclear matter 
phenomenologically and microscopically.  Thanks to developments of neutron 
star observations and nuclear experiments, our interest in the EOS of 
nuclear matter has to extend for a very large region of density and neutron 
excess.  In fact, nuclear matter associated with various systems, e.g., 
stable nuclei, unstable nuclei, nuclear pasta, neutron stars, supernova cores,
heavy-ion collisions at intermediate energies, etc., has different density 
and neutron excess.  On the other hand, our understanding is far from 
sufficient.  Relatively well-known are pure neutron matter, which has
been recently investigated from chiral effective theory interactions 
\cite{EFT}, and symmetric nuclear matter near normal nuclear density,
which reflects the saturation of the nuclear binding energy and density.  
From there, theoretical extrapolations are more or less required.  Moreover, 
one may still ask how large the saturation density of symmetric nuclear 
matter is within five percent errors.  We remark that thermal effects 
are also important for supernova cores and heavy-ion collisions.

In describing the energy of uniform nuclear matter as function of density 
$n$ and neutron excess $\alpha=1-2x$ with the proton fraction $x$, 
it is convenient to use an expansion of the energy per nucleon $w$ around the 
saturation point of symmetric nuclear matter \cite{L},
\begin{equation}
    w=w_0+\frac{K_0}{18n_0^2}(n-n_0)^2+ \left[S_0+\frac{L}{3n_0}(n-n_0)
      \right]\alpha^2.
\label{eos0}
\end{equation}
The parameters characterizing this expansion include the saturation density 
$n_0$ and energy $w_0$ of symmetric nuclear matter, the symmetry energy 
coefficient $S_0$, the incompressibility of symmetric nuclear matter $K_0$, 
and the density symmetry coefficient $L$.  The parameters $L$ and $S_0$ are 
associated with the density dependent symmetry energy coefficient $S(n)$ as
$S_0=S(n_0)$ and $L=3n_0(dS/dn)_{n=n_0}$.  Basically, the parameter $L$ 
corresponds to the pressure of pure neutron matter at $n=n_0$.  Generally, 
higher order coefficients with respect to density such as $K_0$ and $L$ are 
more difficult to determine.  We remark that expression (\ref{eos0})
does not contain even higher order terms, one of which is associated with
the isospin dependence of the incompressibility.

From the viewpoint of microscopic calculations, even pure neutron matter 
at low densities is not simple .  This is because of strong coupling effects 
and uncertainties in the nuclear force.  In fact, the Lee-Yang low density 
expansion only works at very low densities, while we can see a behavior close 
to the unitarity limit 
at densities where the scattering length is very large compared 
with interparticle spacing, which is in turn far larger than the range of the 
interaction \cite{CMPR}.  Fortunately, in addition to variational 
calculations \cite{FP}, elaborate Green's function Monte Carlo calculations 
\cite{CMPR} are 
available for pure neutron matter below normal nuclear density, and they are 
consistent with each other.  More recently, systematic many-body calculations 
based on chiral effective field theory have been performed \cite{EFT}.  
In these calculations, effects of three-body interactions play a role in 
determining the high density behavior of neutron matter EOS.  On the other 
hand, symmetric nuclear matter is still more elusive.  In fact, the saturation 
properties cannot be reproduced by variational calculations without a 
phenomenological three-body force \cite{FP}.  This is partly due to complexity 
involving a strong tensor force.

In this article, we focus on a phenomenological approach to the EOS of 
nuclear matter.  In sect.\ 2, a macroscopic nuclear model, which is
constructed in such a way as to depend on the EOS of uniform nuclear matter, 
is reviewed.  We show that $L$ and $K_0$ remain uncertain while empirical 
masses and charge radii of stable nuclei are equally well reproduced.  
The nuclear model is then used to describe matter radii and masses of 
unstable nuclei, proton elastic scattering and total reaction cross sections 
off unstable nuclei, and nuclei in neutron star crusts including nuclear 
pasta, which are given in sect.\ 3--6.  In sect.\ 7, possible constraints on 
the density dependence of the symmetry energy from experiments on unstable 
nuclei and observations of quasi-periodic oscillations (QPOs) in giant flares 
of soft gamma-ray repeaters (SGRs) are discussed.  
Concluding remarks are finally given in sect.\ 8.

\section{Macroscopic nuclear model}
\label{models}

     From now on, we will focus on a phenomenological approach to the EOS of
nuclear matter.  First we discuss the static properties of atomic nuclei and 
their relation with the EOS of nuclear matter on the basis of ref.\ 
\cite{OI03}.  
To this end, we describe a macroscopic nuclear model in a manner that depends
on the EOS of nuclear matter.  Various applications can arise therefrom.
For example, we will address in the next section how one can extract the 
saturation properties of asymmetric nuclear matter from the size of unstable 
nuclei.  The main conclusion will be that it might be possible that we 
determine the density dependence of the symmetry energy from future 
systematic data for radii of unstable nuclei.  

     Application to neutron stars is of great significance because the EOS of 
nuclear matter at large neutron excess is relevant to the structure and 
evolution of neutron stars \cite{LP04}, 
which are expected to be further clarified by 
future space and ground-based observations.  In the outer part (crust) of a 
star, nuclei present are considered to be very neutron rich or even drip 
neutrons in the presence of a neutralizing background of electrons.  Near 
normal nuclear density, the system is considered to melt into uniform nuclear 
matter.  This nuclear matter mainly constitutes the star's core and thus
controls the structure of a neutron star.

    The symmetry energy is related to the structure and evolution of neutron 
stars in many respects.  
%For example, the crust-core boundary is affected by 
%the symmetry energy.  I will come to that later.
%%%%%%%%%%%%%%%%%%%%%%%%%%%%%%%%%%%%%%%%%%%%%%%%%%%%%%%%%%%%%%%%%%%%%%%%
For example, the mass and radius of a neutron star are mainly determined by 
the EOS of uniform nuclear matter.  The symmetry energy acts to stiffen the 
EOS in a neutron-rich situation as encountered in a star.  Also, neutron 
star cooling is related to the symmetry energy since it controls the proton 
fraction in the core region.  Fast neutrino emission process, i.e., direct 
URCA, can only occur at relatively high proton fraction. 
%%%%%%%%%%%%%%%%%%%%%%%%%%%%%%%%%%%%%%%%%%%%%%%%%%%%%%%%%%%%%%%%%%%%%%%%

  Let us now start with the phenomenological expression for the energy of 
nuclear matter having the neutron and proton number densities $n_n$ and $n_p$, 
which is divided into the kinetic energy part and the potential energy part 
as \cite{O}
\begin{equation}
  w=\frac{3 \hbar^2 (3\pi^2)^{2/3}}{10m_n n}(n_n^{5/3}+n_p^{5/3})
      +(1-\alpha^2)\frac{v_s(n)}{n}+\alpha^2 \frac{v_n(n)}{n},
\label{eos1}
\end{equation}
where
\begin{equation}
  v_s=a_1 n^2 +\frac{a_2 n^3}{1+a_3 n}
\label{vs}
\end{equation}
and
\begin{equation}
  v_n=b_1 n^2 +\frac{b_2 n^3}{1+b_3 n}
\label{vn}
\end{equation}
are the potential energy densities for symmetric nuclear matter and pure
neutron matter, and $m_n$ is the neutron mass (for simplicity we here identify
the proton mass $m_p$ with $m_n$).  The parameter $b_3$, which controls 
the EOS of matter for large neutron excess and high density and thus has 
little effect on the saturation properties of nearly symmetric nuclear matter,
is set to 1.58632 fm$^3$, which was obtained by one of the 
authors \cite{O} in such a way as to reproduce the neutron matter energy of 
Friedman and Pandharipande \cite{FP}. 
In the present energy expression, the potential energy part is a parabolic 
function of $\alpha$, while the kinetic energy part includes higher order 
terms in $\alpha$.  Such $\alpha$ dependence of the potential energy part
is partially justified by variational calculations of Lagaris and 
Pandparihande \cite{LP}.

     Expression (\ref{eos1}) is one of the simplest that reduces to the usual 
expression (\ref{eos0}) near the saturation point of symmetric nuclear matter.
From empirical masses and radii of stable 
nuclei, as we shall see, one can well determine the saturation density $n_0$
and energy $w_0$ of symmetric nuclear matter and the symmetry energy 
coefficient $S_0$ \cite{OI03}.  The incompressibility $K_0$ and the density 
symmetry coefficient $L$ are relatively uncertain, but they control the 
saturation points at finite neutron excess.  In fact, as the neutron excess 
increases from zero, the saturation point moves in the density versus energy 
plane.  Up to second order in $\alpha$, the saturation energy $w_s$ and 
density $n_s$ are given by
\begin{equation}
  w_s=w_0+S_0 \alpha^2
\label{ws}
\end{equation}
and
\begin{equation}
  n_s=n_0-\frac{3 n_0 L}{K_0}\alpha^2.
\label{ns}
\end{equation}
The slope, $y$, of the saturation line near $\alpha=0$ $(x=1/2)$ is thus
expressed as
\begin{equation}
 y=-\frac{K_0 S_0}{3 n_0 L}.
\label{slope}
\end{equation}

\begin{figure*}
\resizebox{0.95\textwidth}{!}{%
  \includegraphics{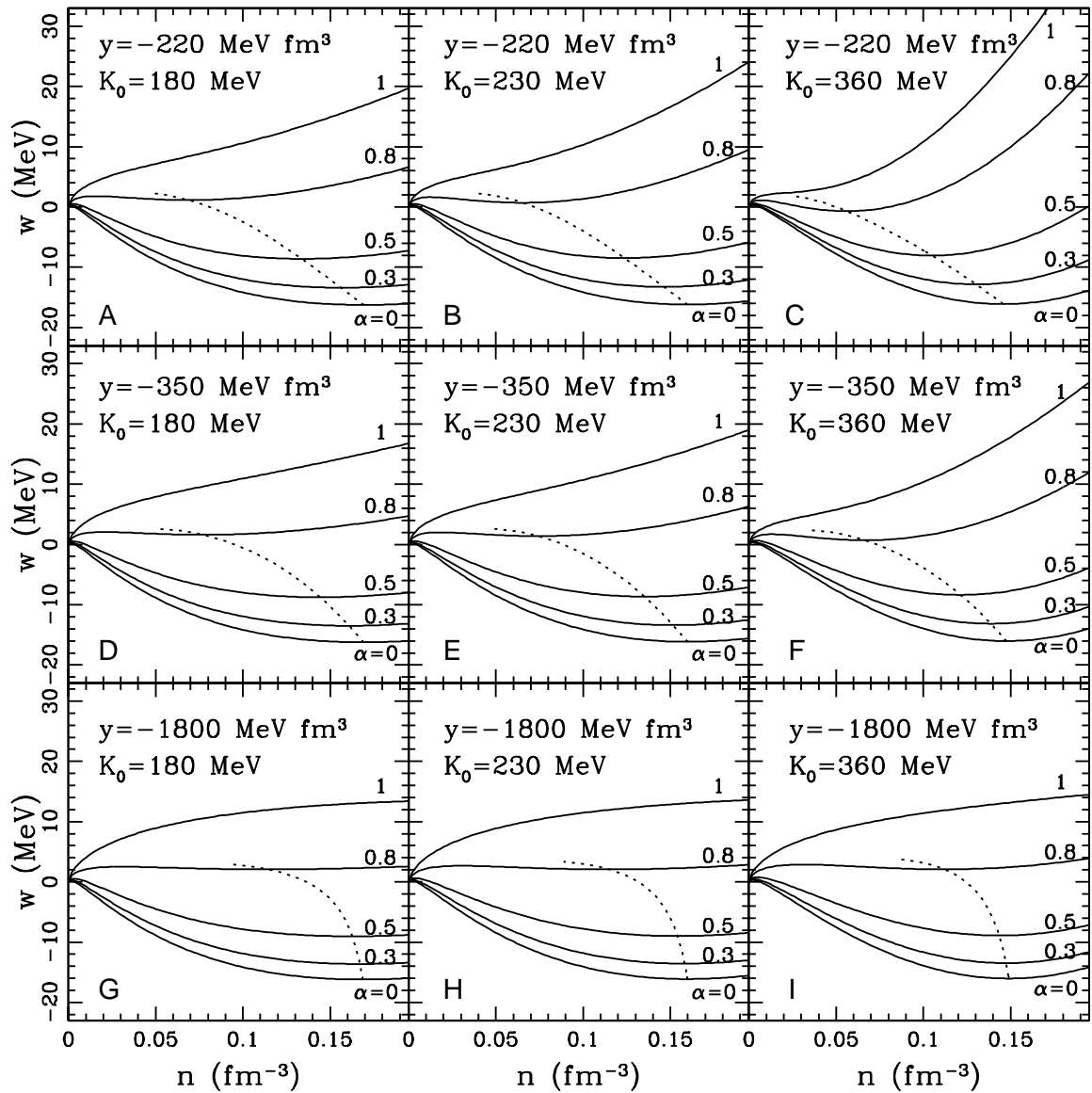}
}
%\vspace*{5cm}       % Give the correct figure height in cm
\caption{Energy per nucleon of nuclear matter for nine extreme 
cases.  In each panel, the solid lines are the energy at neutron
excess $\alpha=0, 0.3, 0.5, 0.8, 1$, and the dotted line is the
saturation line. From ref.\ \cite{OI07}.}
\label{eos}       % Give a unique label
\end{figure*}

   Figure \ref{eos} illustrates nine of the present EOS models, which can be 
obtained for various sets of the incompressibility $K_0$ and the density 
symmetry coefficient $L$ as will be shown below.  
In each panel we plot the energy as function of nucleon 
density ranging from symmetric nuclear matter to pure neutron matter.  The 
saturation line is written in dashed line.  As the incompressibility $K_0$ 
increases, the curvature at the saturation point becomes larger.  While, as 
the density symmetry coefficient $L$ increases, the slope of the saturation 
line becomes gentler.  The question of interest here is what 
kind of EOS is favored by empirical data on nuclear masses and radii.

    For this purpose, we describe macroscopic nuclear properties in a way 
that is dependent on the EOS of nuclear matter by using a simplified version 
of the Thomas-Fermi model \cite{OI03}.  A similar approach was independently
utilized by Bodmer and Usmani \cite{BU}.  The essential point of the present 
model is to
write down the binding energy of a nucleus of mass number $A$ and charge 
number $Z$ in a density functional form:
\begin{equation}
 E=E_b+E_g+E_C+Nm_n c^2+Zm_p c^2,
\label{e}
\end{equation}
where 
\begin{equation}
  E_b=\int d^3 r n({\bf r})w\left[n_n({\bf r}),n_p({\bf r})\right]
\label{eb}
\end{equation}
is the bulk energy,
\begin{equation}
  E_g=F_0 \int d^3 r |\nabla n({\bf r})|^2
\label{eg}
\end{equation}
is the gradient energy with adjustable constant $F_0$,
\begin{equation}
  E_C=\frac{e^2}{2}\int d^3 r \int  d^3 r' 
      \frac{n_p({\bf r})n_p({\bf r'})}{|{\bf r}-{\bf r'}|}
\label{ec}
\end{equation}
is the Coulomb energy, and $N=A-Z$ is the neutron number.  This form allows 
us to connect the EOS and the binding energy through the bulk energy part.  
For simplicity we use the following parametrization for the nucleon 
distributions ($i=n,p$):
\begin{equation}
  n_i(r)=\left\{ \begin{array}{lll}
  n_i^{\rm in}\left[1-\left(\displaystyle{\frac{r}{R_i}}\right)^{t_i}\right]^3,
         & \mbox{$r<R_i,$} \\
             \\
         0,
         & \mbox{$r\geq R_i.$}
 \end{array} \right.
\label{ni}
\end{equation}
Here, $R_i$ roughly represents the nucleon radius, $t_i$ the relative surface 
diffuseness, and $n_i^{\rm in}$ the central number density.  The proton 
distribution of the form (\ref{ni}) can fairly accurately reproduce the 
empirical behavior from electron elastic scattering off stable nuclei.

     In order to construct the nuclear model in such a way as to reproduce 
empirical masses and radii of stable nuclei, we first extremized the binding 
energy with respect to the particle distributions for fixed mass number $A$, 
five EOS parameters ($n_0$, $w_0$, $K_0$, $S_0$, and $L$), and gradient 
coefficient $F_0$.  Next, for various sets of the incompressibility $K_0$ and 
the density symmetry coefficient $L$, we obtained the remaining three EOS 
parameters as well as $F_0$ by fitting the calculated optimal values of the 
nuclear charge number, mass, and root-mean-square (rms) charge radius to 
empirical data for stable nuclei on the smoothed beta stability line.  Here 
we have defined the rms charge radius as
\begin{equation}
    R_c=\left[Z^{-1}\int d^3 r r^2 \rho_c({\bf r})\right]^{1/2},
\label{Rc}
\end{equation}
where 
\begin{equation}
    \rho_c({\bf r})=(\pi^{1/2}a_p)^{-3}\int d^3 r' 
                  \exp\left({-|{\bf r}-{\bf r'}|^2/a_p^2}\right)n_p({\bf r'})
\label{rhoc}
\end{equation}
with $a_p=0.65$ fm represents the charge distribution folded with the proton
form factor \cite{ES}.
%%%%%%%%%%%%%%%%%%%%%%%%%%%%%%%%%%%%%%%%%%%%%%%%%%%%%%%%%%%%%%%%%%%%%%%%%%%
      The rms deviations of the calculated masses from the measured values 
\cite{Audi} are of order 3 MeV, which are comparable with the deviations 
obtained from the Weizs{\" a}cker-Bethe mass formula, while the rms deviations 
of the calculated charge radii from the measured values \cite{VJV} are about 
0.06 fm, which are comparable with the deviations obtained from the 
$A^{1/3}$ law.
%%%%%%%%%%%%%%%%%%%%%%%%%%%%%%%%%%%%%%%%%%%%%%%%%%%%%%%%%%%%%%%%%%%%%%%%%%%

%12.  This is a typical example of the optimal particle distributions 
%calculated for Zr and Pb.  The proton distributions agree well with the 
%empirical data from electron elastic scattering.  

\begin{figure}
\resizebox{0.5\textwidth}{!}{%
  \includegraphics{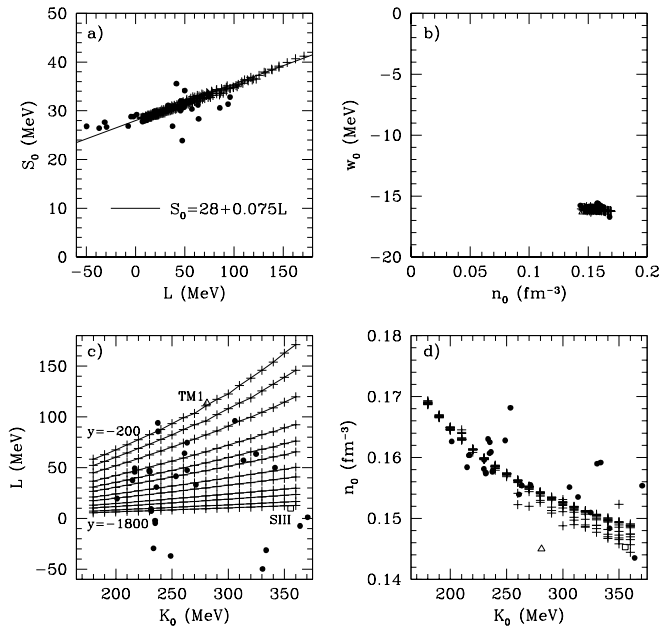}
}
%\vspace{5cm}       % Give the correct figure height in cm
\caption{Various optimal relations among the parameters $S_0$, $n_0$, $w_0$, 
$L$, and $K_0$ characterizing the EOS of nearly symmetric nuclear matter.  In
addition to the present results (crosses), the Skyrme-Hartree-Fock predictions
[dots except for SIII (square)] and the TM1 prediction (triangle) are plotted.
In (c), the thin lines are lines of constant $y$.  From ref.\ \cite{OI03}.
}
\label{eospara}       % Give a unique label
\end{figure}

   In fig.\ \ref{eospara} we exhibit the EOS parameter region that can be 
constrained from the fitting to empirical masses and radii of stable nuclei, 
together with various mean-field-model predictions.  The saturation density 
$n_0$ and energy $w_0$ and the symmetry energy coefficient $S_0$ are fairly 
well constrained, while about 200 sets of the incompressibility $K_0$ and 
the density symmetry coefficient $L$ can provide reasonable fitting.  In fact, 
the data fitting based on the least squares method depends on how to assign 
weights, which makes it impractical to find the optimal values of $K_0$
and $L$ among the sets, while for $y$ higher than $\sim -200$ MeV fm$^3$, 
the fitting becomes no longer effective.  This 
$K_0$-$L$ region is a starting point of our study.  We want to narrow this 
region by using future experiments for unstable nuclei.  In the present 
analysis we rule out the possibility that 
the slope $y$ is positive.  A positive $y$ is inconsistent with the fact that 
the empirical matter radii of $A=17,20,31$ isobars \cite{A17,A20,A31} tend to 
increase with neutron/proton excess.  This is because a positive $y$ plays a 
role in increasing the saturation density $n_s$ with neutron/proton excess, 
as can be seen from eqs.\ (\ref{ns}) and (\ref{slope}).

     It is also interesting to see the roughly linear relation between $L$ 
and $S_0$ in fig.\ \ref{eospara}.  This tendency is not clear in the 
Skyrme-model calculations with zero-range force, but is known to be seen in 
the Gogny-model calculations with finite-range force \cite{FPR}.  How this
tendency is related to the range of the three-nucleon force has been recently 
discussed \cite{GCR}.  We remark, however, that the roughly 
linear relation is consistent with a recent $1\sigma$ fit to experimental 
masses and radii using a Skyrme-parametrized energy density functional 
\cite{Kortelainen10}.

    The present $L$-$S_0$ relation can be compared with other constraints on 
$L$ and $S_0$.  It turns out that various constraints from heavy-ion data 
associated with isospin diffusion and neutron-proton ratio, pigmy dipole 
resonances, excitation energies of isobaric analog states are not always 
consistent with each other \cite{Tsang}.  Something has to be wrong, but we do 
not know which one.  This may imply that data from stable nuclei are not 
enough to reasonably constrain $L$.

\begin{figure}
\resizebox{0.5\textwidth}{!}{%
  \includegraphics{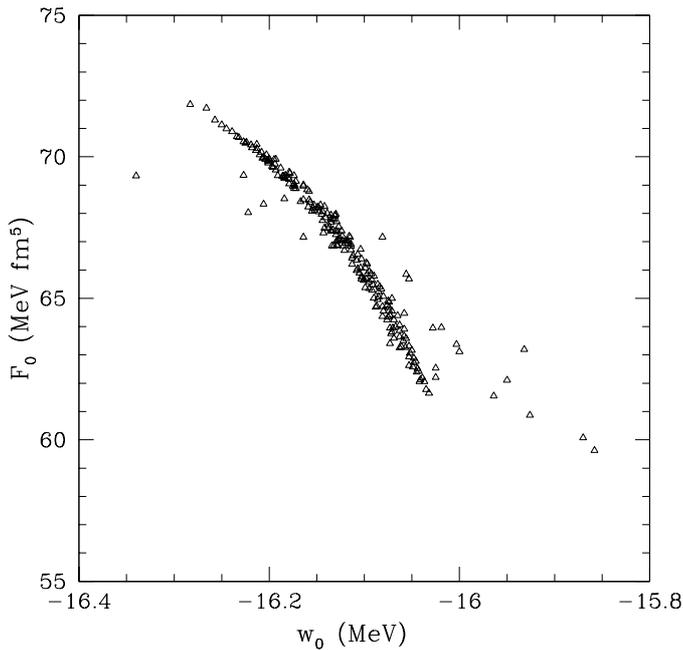}
}
%\vspace{5cm}       % Give the correct figure height in cm
\caption{Optimal relation of the gradient coefficient $F_0$ with 
the saturation energy $w_0$ of symmetric nuclear matter.  
}
\label{f0}       % Give a unique label
\end{figure}

    For completeness, in fig.\ \ref{f0} we exhibit the optimal values of 
$F_0$, which ranges $\sim60$--70 MeV fm$^5$, as a function of the optimal 
$w_0$.  We note that there is a clear correlation 
between $F_0$ and $w_0$ because fitting to empirical masses requires a larger 
gradient energy for a larger bulk binding energy.  

    We conclude this section by summarizing salient features of the 
macroscopic nuclear model used here.  This model can describe global nuclear 
properties such as masses and rms radii in a manner that is dependent on the 
EOS of nuclear matter, although it is not good at describing tails of the 
density distribution and does not allow for shell or pairing effects.  
As will be shown in the next section, the present macroscopic approach 
predicts that the matter radii depend appreciably on the density symmetry 
coefficient $L$, while being almost independent of the incompressibility 
$K_0$.

\section{Matter radii of unstable nuclei}
\label{radii}

\begin{figure*}
\resizebox{0.95\textwidth}{!}{%
  \includegraphics{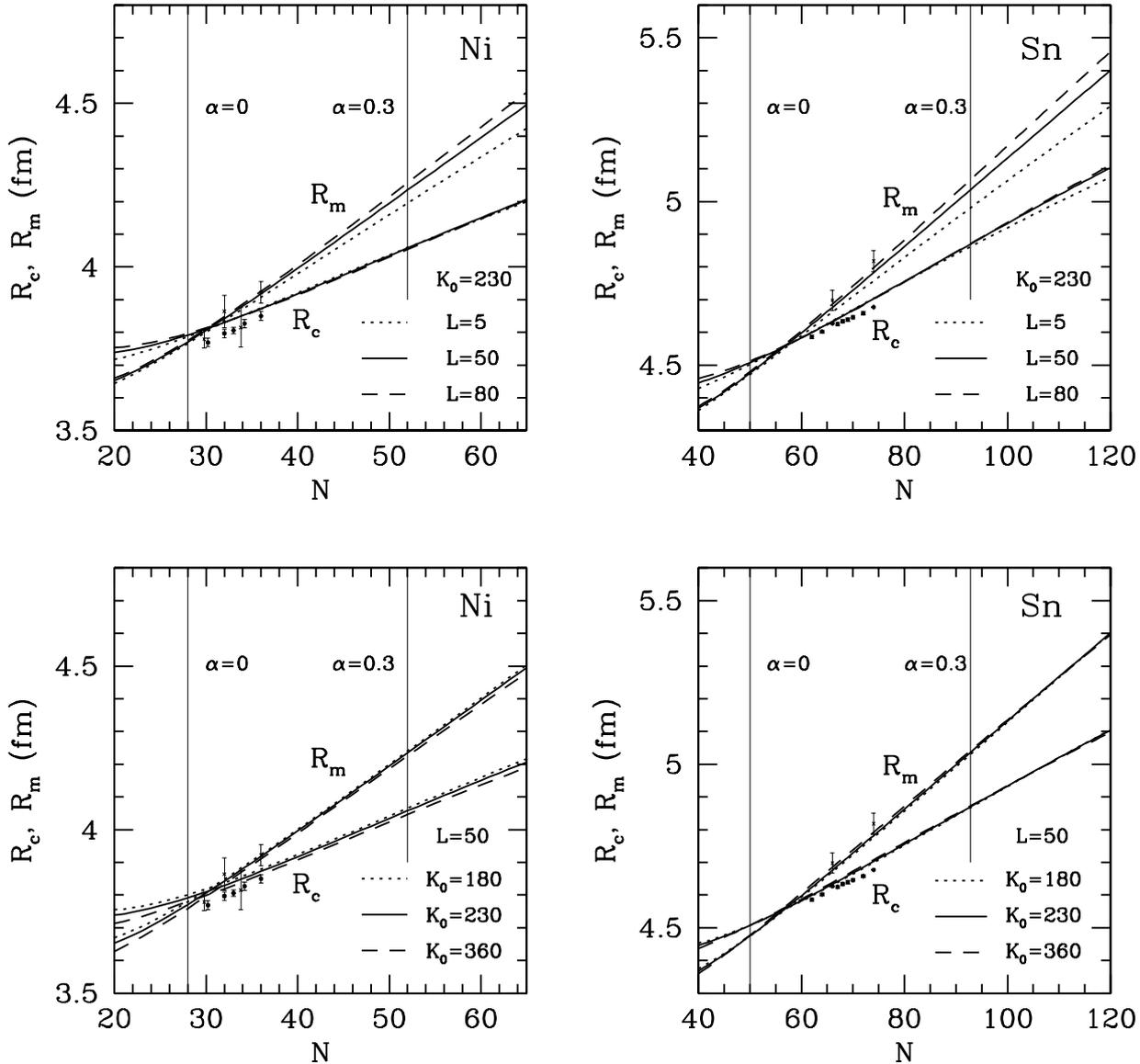}
}
%\vspace*{5cm}       % Give the correct figure height in cm
\caption{
The rms charge and matter radii, $R_c$ and $R_m$,
of Ni and Sn isotopes
for combinations of $L=0,50,80$ MeV and $K_0=180,230,360$
MeV.  The experimental data for the rms charge radii (dots) and
matter radii (crosses) are taken from refs.\ \cite{VJV} 
and \cite{Batty},
respectively.  From ref.\ \cite{OI03}.}
\label{size}       % Give a unique label
\end{figure*}

    We now address how the EOS-dependent macroscopic nuclear model predicts 
the rms matter radii of unstable nuclei.  As clarified in the previous section,
the saturation density $n_0$ and energy $w_0$ and the symmetry energy 
coefficient $S_0$ can be well determined from systematic data on masses and 
radii of stable nuclei, while the incompressibility $K_0$ and the density 
symmetry coefficient $L$ are more difficult to determine.  There are many 
trials of extracting the incompressibility from empirical data such as giant 
monopole resonances in stable nuclei \cite{JP} and even caloric curves in 
nuclear collisions \cite{Natowitz}.  Unfortunately this kind of extraction 
depends on models for 
the effective nucleon-nucleon interaction \cite{KM}.  A decade ago, we 
proposed a method for extracting the density derivative of the symmetry energy 
from future systematic data on the matter radii of unstable nuclei on the 
basis of the macroscopic nuclear model \cite{OI03}.  In RIKEN and GSI, it is 
expected that RI beams of heavy nuclides incident on proton targets will 
provide elastic scattering data with reasonable accuracy, from which one may be
able to deduce the matter radii, e.g., through an empirical relation between 
the first diffraction peak angles and the matter radii \cite{BS1}.  In GSI, 
this type of experiment named S272 was performed for $^{70}$Ni several years 
ago, but the data remain unpublished.

     Figure \ref{size} shows the rms matter and charge radii for Ni and Sn
isotopes calculated for various sets of $K_0$ and $L$. Here we have 
defined the rms matter radius as
\begin{equation}
    R_m=\left[A^{-1}\int d^3 r r^2 \rho_m({\bf r})\right]^{1/2},
\label{Rm}
\end{equation}
where
\begin{equation}
    \rho_m({\bf r})=(\pi^{1/2}a_p)^{-3}\int d^3 r' 
                     \exp\left({-|{\bf r}-{\bf r'}|^2/a_p^2}\right)n({\bf r'})
\label{rhom}
\end{equation}
is the matter distribution folded with the proton charge form factor equally
for neutrons and protons.  At fixed $K_0$, 
differences of order 0.1 fm occur in the prediction of the matter radii of 
very neutron-rich nuclei due to uncertainties in the density symmetry 
coefficient $L$.  This tendency arises because the saturation density at 
nonzero neutron excess decreases with increasing $L$ as in eq.\ (\ref{ns}).  
On the other hand, the matter radii are almost independent of the 
incompressibility $K_0$.  Note that as $K_0$ increases, the surface 
diffuseness is reduced, while the saturation density $n_0$ is also reduced as 
shown in fig.\ \ref{eospara}(d).  We can thus conclude that these effects 
counteract with each other.  Such $K_0$ independence is promising for the 
purpose of deriving the value of $L$ from the experimentally deduced matter 
radii.  These are just plotted for stable nuclei by crosses, which are 
deduced from proton elastic scattering data by using optical potential 
models, but are not useful for derivation of $L$.  
Data for unstable nuclei are thus strongly desired.

    It is often claimed that the neutron skin thickness of neutron-rich stable
nuclei such as $^{124}$Sn and $^{208}$Pb can severely constrain $L$ (e.g., 
ref.\ \cite{RMCVW}).  However,
theoretical predictions of the neutron skin thickness depend not only on the 
nuclear bulk properties but also on the nuclear surface properties.  In fact, 
within a compressible liquid-drop model \cite{Y,IO}, one can show that the 
predicted neutron skin thickness, which is determined by a balance between
the bulk and surface symmetry energies, has a linear dependence on $L$, but the
poorly known density dependence of the surface tension prevents a 
model-independent constraint on $L$.  Although the macroscopic nuclear model
used here and various mean-field models give a roughly linear correlation 
between $L$ and the neutron skin thickness, respectively, such a correlation 
is significantly different between the two types of models.  This kind of 
difference would affect a possible constraint on $L$.  Note that
the present macroscopic model tends to underestimate the surface diffuseness, 
which is not favorable for the prediction of the neutron skin thickness.

    Several summarizing remarks on the contents in this and previous sections 
are in order.  By using the macroscopic nuclear model, we derived the 
relations between the EOS parameters from experimental data on the masses and 
charge radii of stable nuclei, and we found that $L$ and $K_0$ are still 
uncertain.  The important prediction is that the density symmetry coefficient 
$L$ may be determined if a global behavior of the matter radii at large neutron
excess is obtained from future systematic measurements of the matter radii of 
unstable nuclei.  Lastly, we remark that the parameter $L$, which characterizes
the dependence of the EOS on neutron excess, is relevant to the structure and 
evolution of neutron stars through mass-radius relation, crust-core boundary, 
cooling, etc.

\section{Proton elastic scattering and total reaction cross sections off 
unstable nuclei}
\label{exp}

     It is not straightforward to deduce the matter radii of unstable nuclei
from experimental data such as proton-nucleus elastic differential cross
sections and total reaction cross sections because it requires the approximate 
scattering theory \cite{Batty,Ray}.  It is thus instructive to examine how 
the cross sections themselves are related to the parameter $L$ in a proper 
theoretical framework, namely, the optical limit approximation of the Glauber 
multiple scattering model \cite{G} that incorporates the nucleon distributions 
as obtained in sect.\ 2.

    We first consider proton elastic scattering on the basis of ref.\ 
\cite{IOB}.  For sufficiently high proton
incident energies and small momentum transfers to validate the Glauber model, 
we find that the angle of the scattering peak decreases with $L$ more 
remarkably for larger neutron excess, while the peak height increases with 
$K_0$ almost independently of neutron excess.  We suggest the possibility that 
comparison of the calculations with experimental data for the peak angle may be
useful for determination of $L$.

    The elastic differential cross section at given momentum 
transfer ${\bf q}$ and incident proton energy $T_p$ can be written as (e.g., 
ref.\ \cite{Ahmad})
\begin{equation}
\frac{d\sigma}{d\Omega}=|F({\bf q})|^2,
\end{equation}
with the elastic scattering amplitude,
\begin{equation}
|F({\bf q})|=|F_{\rm C}({\bf q})+\frac{ik}{2\pi}\int d{\bf b}
e^{-i{\bf q\cdot b}+2i\eta\ln(k|{\bf b}|)}
\left[1-e^{i\chi_{\rm N}({\bf b})}\right]|.
\label{amp}
\end{equation}
%\begin{equation}
%|F({\bf q})|=\left|F_{\rm C}({\bf q})+\frac{ik}{2\pi}\int d{\bf b}
%e^{-i{\bf q\cdot b}+2i\eta\ln(k|{\bf b}|)}
%\left[1-e^{i\chi_{\rm N}({\bf b})}\right]\right|.
%\label{amp}
%\end{equation}
Here, $\hbar k=\sqrt{(T_p/c+m_p c)^2-(m_p c)^2}$ is the incident proton 
momentum, ${\bf b}$ is the impact parameter, $\eta=Ze^2/\hbar v$ with the 
incident proton velocity $v=\hbar k c/(T_p/c+m_p c)$ is the Sommerfeld 
parameter, 
\begin{equation}
F_{\rm C}({\bf q})=-\frac{2\eta k}{{\bf q}^2}
     \exp\left[-2i\eta\ln\left(\frac{|{\bf q}|}{2k}\right)
       +2i {\rm arg}\Gamma(1+i\eta)\right]
\end{equation}
is the amplitude of the Coulomb elastic scattering, which we approximate as 
a usual Rutherford scattering off a point charge, and 
\begin{equation}
i\chi_{\rm N}({\bf b})=-\int d{\bf r} [n_p({\bf r})\Gamma_{pp}({\bf b}-{\bf s})
+n_n({\bf r})\Gamma_{pn}({\bf b}-{\bf s})]
\label{phase}
\end{equation}
is the phase shift function with the projection ${\bf s}$ of the coordinate
${\bf r}$ on a plane perpendicular to the incident proton momentum and with
the profile function $\Gamma_{pN}$ of the free proton-nucleon ($pN$) 
scattering amplitude, for which we use a simple parametrization,
\begin{equation}
\Gamma_{pN}({\bf b})=\frac{1-i\alpha_{pN}}{4\pi\beta_{pN}}\sigma_{pN}
                 \exp(-{\bf b}^2/2\beta_{pN}),
\label{profile}
\end{equation}
where $\alpha_{pN}=-{\rm Im}\Gamma_{pN}(0)/{\rm Re}\Gamma_{pN}(0)$,
$\sigma_{pN}$ is the $pN$ total cross section, and $\beta_{pN}$ is the slope 
parameter.  Here the values of $\alpha_{pN}$, $\beta_{pN}$, and $\sigma_{pN}$ 
at given incident proton energy $T_p$ are taken from ref.\ \cite{Ray1}.

     Generally, the peak angles are related to the nuclear size, while the 
peak heights are related to the surface diffuseness.  In our macroscopic 
calculations, the radius and diffuseness are in turn related to the EOS 
parameters.  For larger density symmetry coefficient $L$ we obtain a larger 
radius, while for larger incompressibility $K_0$ we obtain a smaller surface
diffuseness.  So it is interesting to investigate the detailed peak structure 
in the small angle regime and its relation with the EOS parameters.

\begin{figure}
\resizebox{0.5\textwidth}{!}{%
  \includegraphics{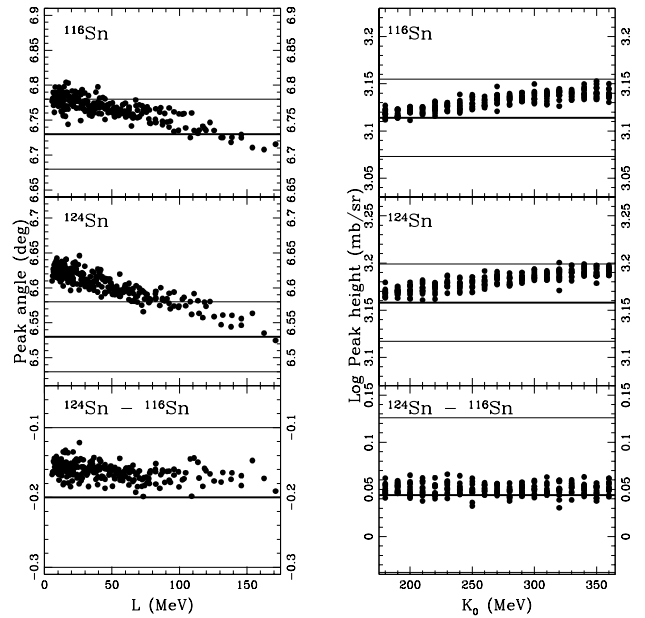}
}
%\vspace{5cm}       % Give the correct figure height in cm
\caption{
The angles and heights of the scattering peak in the small angle
regime, calculated as functions of $L$ and $K_0$ for $p$-$^{116}$Sn and 
$p$-$^{124}$Sn elastic scattering at $T_p=800$ MeV.  The experimental angles 
and heights including errors (from ref.\ \cite{Ray2}) are denoted by the
horizontal lines (thick lines: central values, thin lines: upper and lower
bounds).  From ref.\ \cite{IOB}.
}
\label{peak1}       % Give a unique label
\end{figure}

    In fig.\ \ref{peak1} we illustrate the scattering angles and heights in the
first peak, calculated for about 200 sets of the EOS parameters in the case of 
stable Sn isotopes at incident energy of 800 MeV.  The peak angle decreases 
with $L$.  This is an important 
property which might enable us to extract $L$ from comparison with the 
experimental peak angle.  However, such extract is difficult in this case 
because the experimental uncertainty due to the absolute angle calibration, 
which is taken to be 0.05 deg, is too large to distinguish between different 
$L$'s for nuclei having neutron excess of order or smaller than 0.2.

     On the other hand, the peak height increases with the incompressibility 
$K_0$ in a way almost independent of neutron excess.  However, it is also 
difficult to extract $K_0$ from comparison with the experimental peak
height mainly because our semi-classical nuclear model tends to underestimate 
the surface diffuseness.

\begin{figure}
\resizebox{0.5\textwidth}{!}{%
  \includegraphics{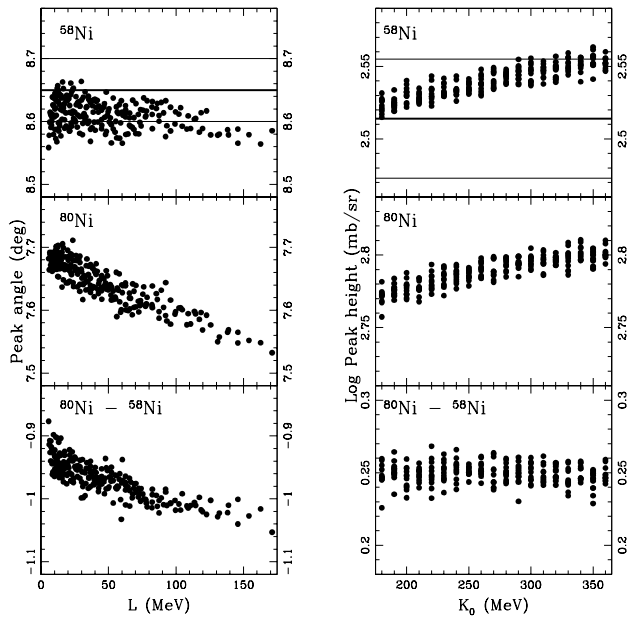}
}
%\vspace{5cm}       % Give the correct figure height in cm
\caption{
Same as fig.\ \ref{peak1} for $p$-$^{58}$Ni and $p$-$^{80}$Ni elastic 
scattering at $T_p=800$ MeV.  From ref.\ \cite{IOB}.
}
\label{peak2}       % Give a unique label
\end{figure}

     We move on to scattering off unstable nuclei, whose beams incident
on proton targets can provide elastic scattering data.  Here we consider a 
very neutron-rich nucleus $^{80}$Ni.  In fact, the neutron excess for 
$^{80}$Ni amounts to 0.3.  We perform the calculations for $^{58}$Ni and 
$^{80}$Ni at incident energy of 800 MeV.  We find from fig.\ \ref{peak2}
that the $L$ dependence of 
the difference in the peak angle between the $^{58}$Ni and $^{80}$Ni cases 
looks large enough to enable us to extract $L$.  It is now useful to take the
difference because our calculations based on the macroscopic nuclear model 
contain systematic errors by ignoring pairing and shell effects and tails 
of the nucleon distributions.  In order to take full care of such systematic 
errors, systematic measurements of the peak angles in the small angle region 
for various nuclides are desired for as large neutron excess as possible with 
accuracy of order 0.01 deg.  According to the usual Fraunhofer diffraction, the
difference of order 0.01 deg in the peak angle corresponds to the difference 
of order 0.01 fm in the rms matter radius.  We remark in passing that the 
present analysis eventually developed into a black sphere model \cite{BS1} 
that gives a nuclear length scale that characterizes measured total reaction 
cross sections and elastic diffraction peak angles simultaneously.

%    Reaction cross section (only for 800 MeV, fig. 1 of IOBK) 
%Glauber perdiction shows only a weak L dependence
%Why [IOBK]

      We finally consider proton-nucleus total reaction cross sections, which 
can be calculated from the same Glauber model as used for the elastic 
scattering calculations \cite{IOBK}.  The total reaction cross section can be 
written as
\begin{equation}
\sigma_R=\int d{\bf b}\left(1-\left|e^{i\chi_{\rm N}({\bf b})}
                               \right|^2\right),
\end{equation}
where $\chi_{\rm N}$ is given by eq.\ (\ref{phase}).

\begin{figure}
\resizebox{0.5\textwidth}{!}{%
  \includegraphics{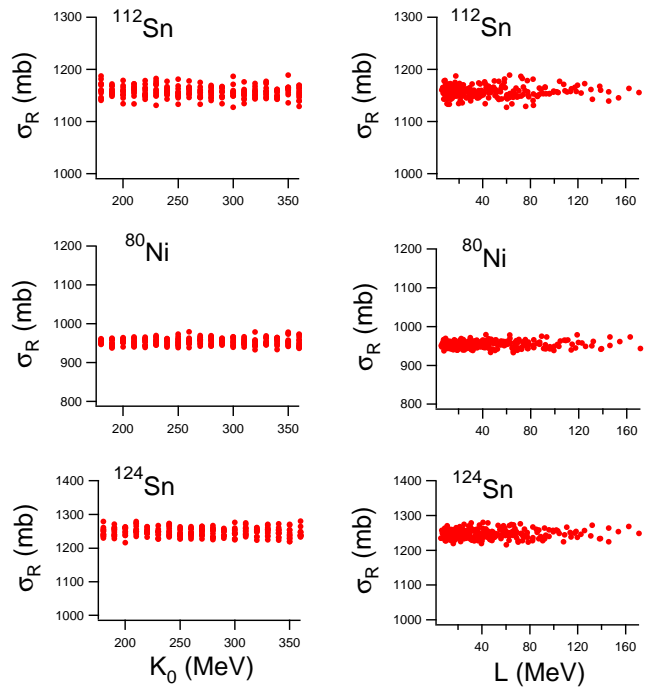}
}
%\vspace{5cm}       % Give the correct figure height in cm
\caption{
(Color online) The total reaction cross sections calculated
as a function of $K_{0}$ and $L$ for $p$-$^{112,124}$Sn 
and $p$-$^{80}$Ni at $T_p=800$ MeV.  From ref.\ \cite{IOBK}.
}
\label{800.reac}       % Give a unique label
\end{figure}

     Figure~\ref{800.reac} shows the results for the selected isotopes 
at $T_p=800$ MeV.  The dependence of the total reaction cross section 
$\sigma_R$ on the EOS parameters is weak even for very neutron-rich nuclei 
such as $^{80}$Ni, in contrast to the case of elastic scattering in which a 
strong $L$ (or size) dependence of the calculated diffraction peak angle 
appears for $^{80}$Ni as shown in fig.\ \ref{peak2}.  This was not expected 
from a standard picture that the larger size, the larger $\sigma_R$, but in 
fact reflects a feature of the optical limit Glauber theory in which an 
unphysical
exponential dependence of the reaction cross section on the neutron skin 
thickness remains when the total proton-neutron cross section is small 
enough \cite{IOBK}.  For duly describing the size dependence of the total 
reaction cross sections, therefore, alternative approaches based on empirical 
data for the total reaction cross sections such as those in refs.\ \cite{IL} 
and \cite{IKO} might be useful even for high incident energies where the 
Glauber theory is usually assumed to be valid.

    We remark that differences between interaction cross sections and total 
reaction cross sections are often ignored, which causes interaction cross 
sections, whose data are far easier to obtain experimentally, to be 
identified with total reaction cross sections.  However, this is not always 
the case even for high energy data, as suggested by using pseudodata for 
total reaction cross sections that can be obtained from the measured elastic 
diffraction peak angles for stable nuclei via the black sphere model 
\cite{KIO08}.
Within the framework of the full  Glauber scattering theory, Novikov and 
Shabelski \cite{NS} also confirmed that the differences are appreciable.

\section{Nuclear masses of unstable nuclei}
\label{mass}

    Let us move on to nuclear masses.  The mass has an advantage over the 
size because empirical mass data have been already accumulated for unstable 
nuclei.  Then, it is natural to ask if the existing data for masses of 
unstable nuclei is useful for determination of $L$.  We shall give a
tentative answer on the basis of refs.\ \cite{OI10,OIK10}.

   We want to know the global neutron excess dependence of nuclear masses.
To this end, some kind of differentials are useful.  Here we focus on the two 
proton separation energy $S_{2p}(Z,N)=E_B(Z,N)-E_B(Z-2,N)$ with the nuclear
binding energy $E_B$.  As illustrated in ref.\ \cite{KTUY}, for fixed proton 
number $Z$, the empirical values of $S_{2p}$ and also the values 
from the Koura-Tachibana-Uno-Yamada (KTUY) 
mass formula show a very smooth isospin dependence except for 
proton shell gaps.  Moreover, the even-odd staggering essentially disappears 
in the two proton separation energy $S_{2p}$.  On the other hand, the two 
neutron separation energy $S_{2n}=E_B(Z,N)-E_B(Z,N-2)$ does not show an ideal 
isospin dependence at fixed $Z$, because it has a very large discontinuity 
around neutron magic numbers.

\begin{figure}
\resizebox{0.5\textwidth}{!}{%
  \includegraphics{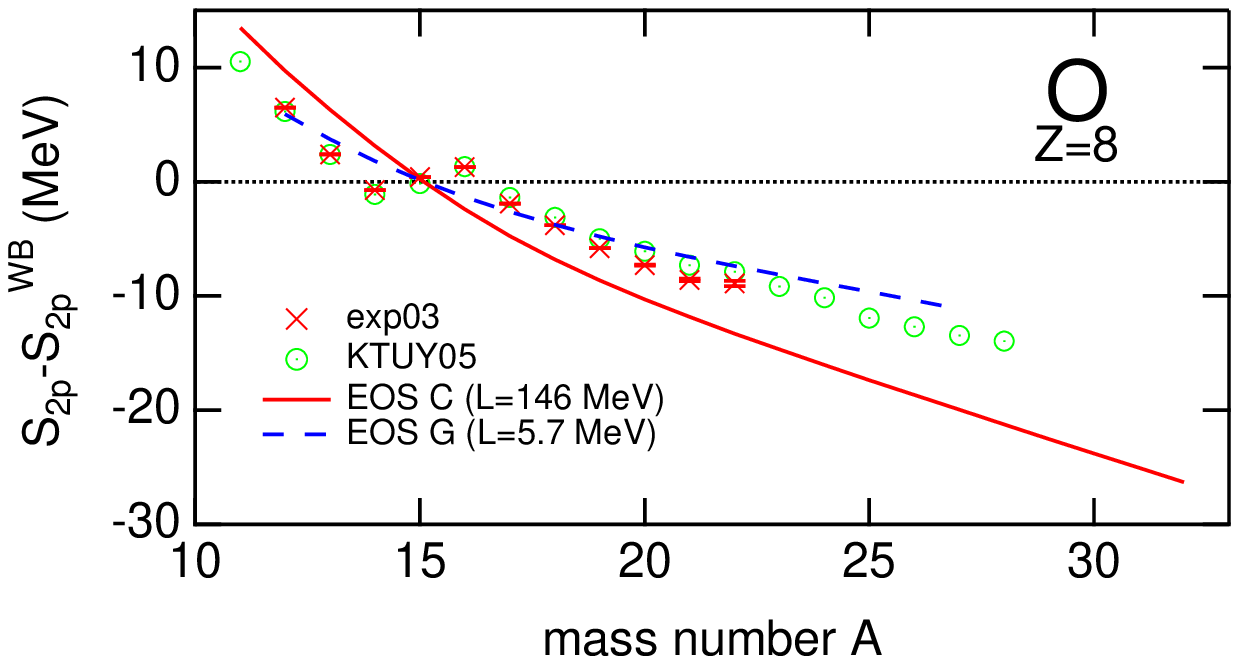}
}
\resizebox{0.5\textwidth}{!}{%
  \includegraphics{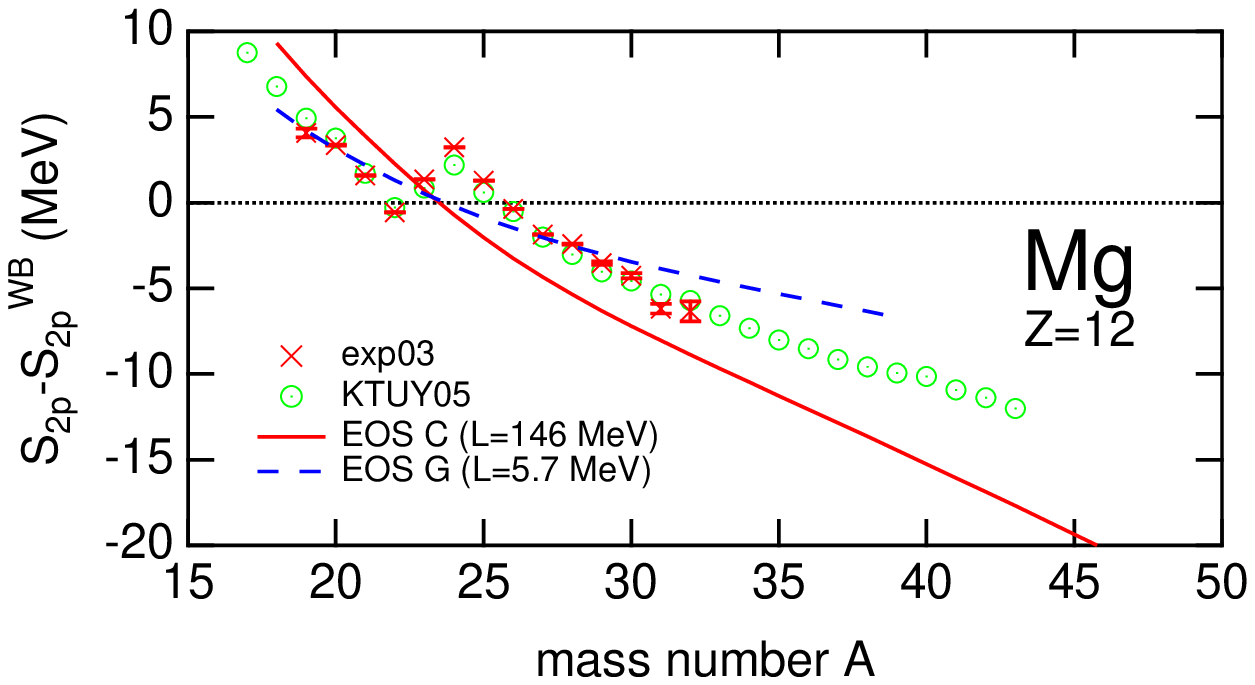}
}
\resizebox{0.5\textwidth}{!}{%
  \includegraphics{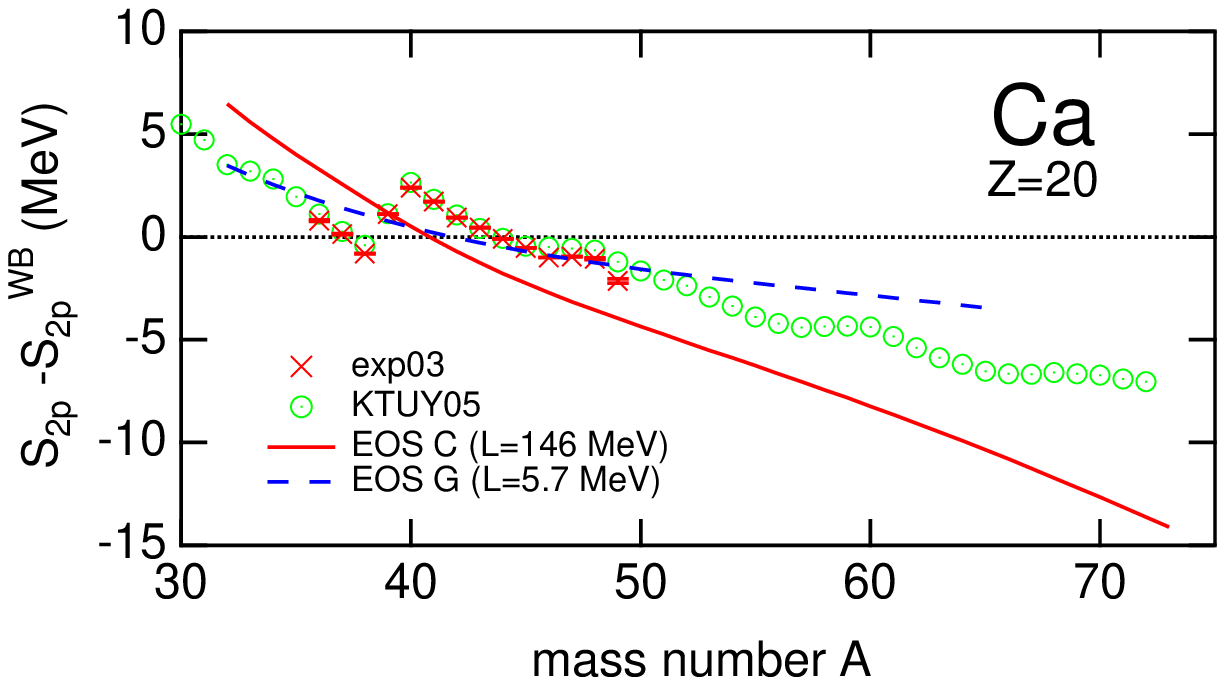}
}
\resizebox{0.5\textwidth}{!}{%
  \includegraphics{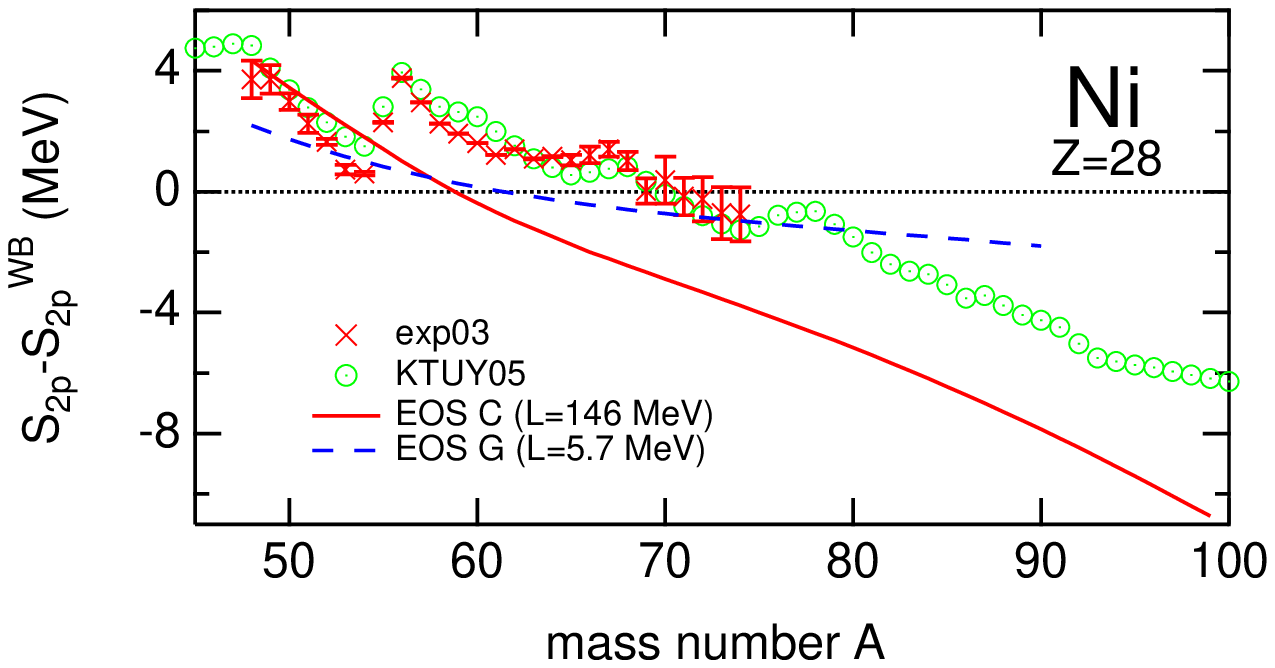}
}
%\vspace{5cm}       % Give the correct figure height in cm
 \caption{(Color online)
The two-proton separation energy, having the one calculated from a
Weizs{\" a}cker-Bethe mass formula \cite{Y} subtracted out, for O, Mg, Ca, 
and Ni
%Sn, and Pb 
isotopes.  The empirical values \cite{Audi}, the calculated values
from the EOS models C and G, and the values obtained from a contemporary mass
formula \cite{KTUY} are plotted in each panel.  From ref.\ \cite{OI10}.}
\label{S2p} 
\end{figure}

    We now try to compare the empirical $S_{2p}$ and the predictions from 
the EOS models C and G shown in fig.\ \ref{eos} by using the macroscopic 
nuclear model.   As exhibited in fig.\ 
\ref{S2p}, the empirical $S_{2p}$ shows a smooth dependence on neutron excess 
except for symmetric nuclei, nuclei with neutron magic numbers, and deformed 
nuclei.  On the other hand, the calculated $S_{2p}$ shows a larger neutron 
excess dependence for larger $L$.  Comparison between the empirical and 
calculated $S_{2p}$ seems easier for smaller mass.  As far as the slope 
of $S_{2p}$ with respect to neutron excess is concerned, a larger 
$L$ value is more consistent with the empirical data.  Note that there are 
roughly uniform offsets between the data and the calculations with
the EOS model C at $N>Z$, which are presumably due to proton shell gaps
ignored in the present calculations.  That is why the fact that 
the calculated $S_{2p}$ from the EOS model G is apparently closer to the
empirical $S_{2p}$ has to be seen with caution.

\begin{figure}
\resizebox{0.5\textwidth}{!}{%
  \includegraphics{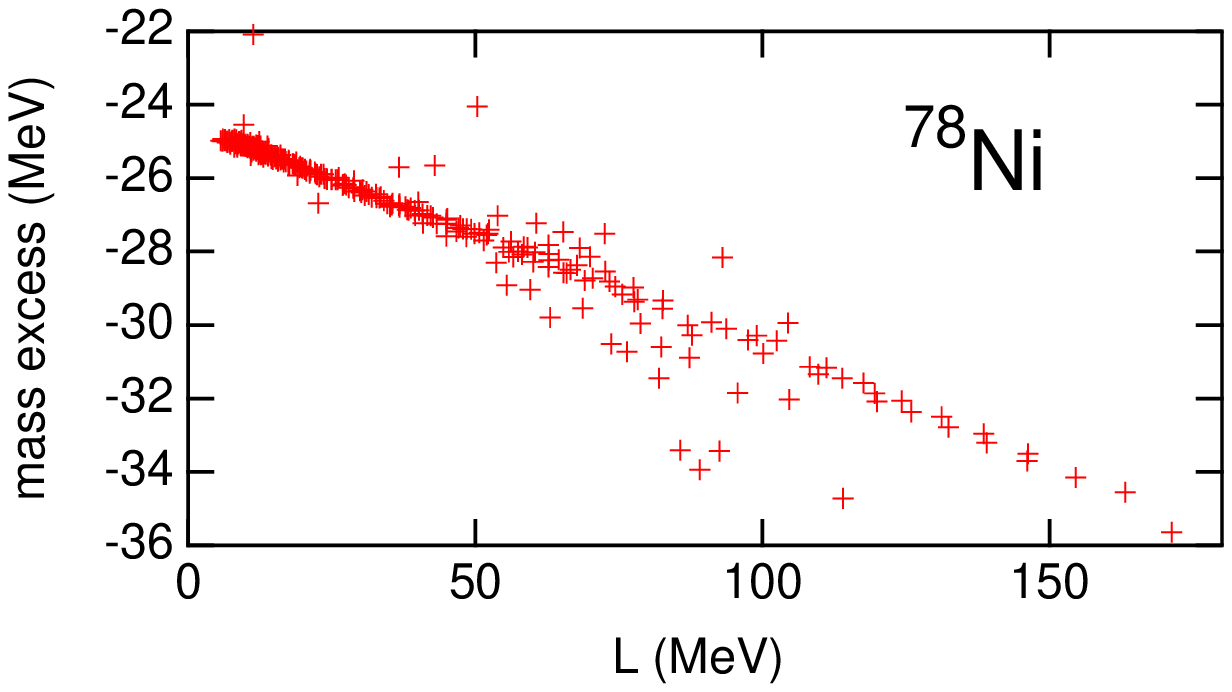}
}
\resizebox{0.5\textwidth}{!}{%
  \includegraphics{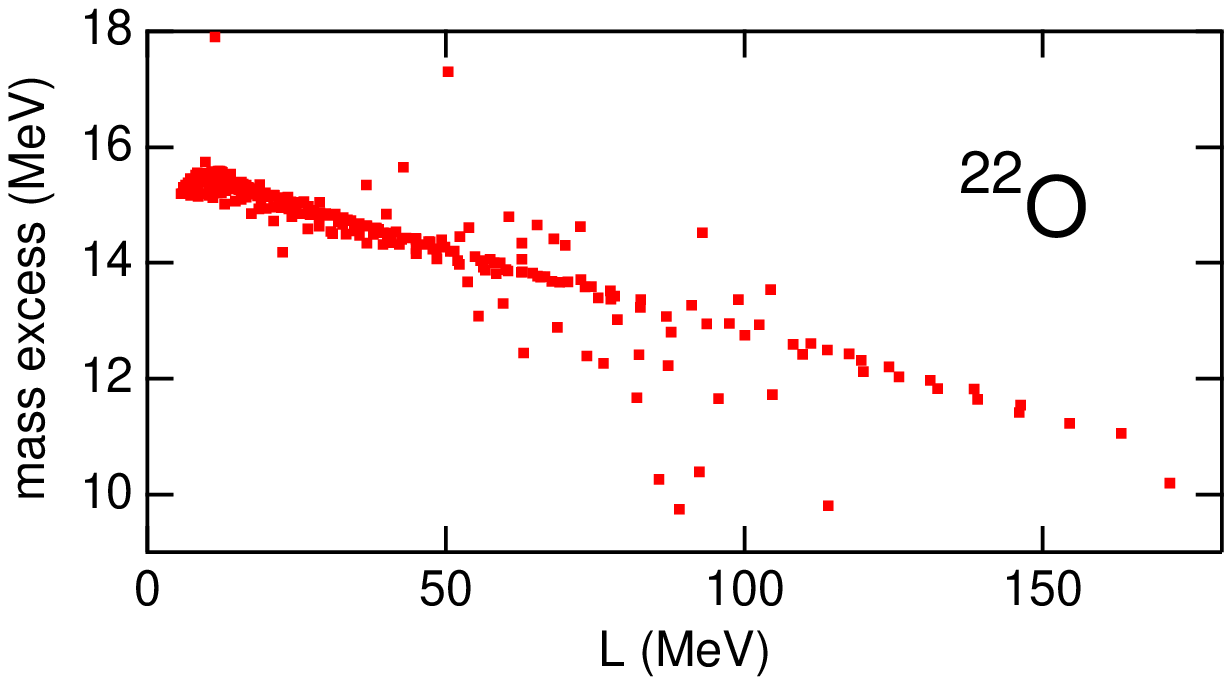}
}
%\vspace{5cm}       % Give the correct figure height in cm
 \caption{(Color online)
The mass excess calculated for $^{78}$Ni and $^{22}$O as a function of $L$.
From ref.\ \cite{OI10}.}
\label{massNi} 
\end{figure}

     In order to understand the $L$ dependence of $S_{2p}$, we go back to the 
$L$ dependence of the calculated nuclear masses.  For very neutron rich nuclei,
as shown in fig.\ \ref{massNi}, the calculated mass decreases with $L$, while 
having a relatively weak dependence on $K_0$.  This suggests that nuclear 
masses are not always dominated by the bulk properties of nuclear matter.  
In fact, the $L$ dependence of the calculated mass cannot be explained by the 
bulk asymmetry term because for a larger $L$ we obtain a larger $S_0$ (see 
fig.\ \ref{eospara}), leading to a larger mass according to eq.\ (\ref{ws}).  
Therefore, we can conclude that the surface asymmetry term is responsible for 
the present $L$ dependence of the calculated mass.  In fact, within a 
compressible liquid-drop model \cite{Y,IO}, one can show that the surface 
tension for neutron rich nuclei is effectively smaller for larger $L$, 
leading to a smaller mass and $S_{2p}$.
 
     Note that the macroscopic nuclear model used here effectively has a
nonvanishing density dependence of the surface tension.  This is a contrast
to the cases of many compressible liquid-drop models 
%in which the Myers-Swiatecki ``theorem'' that claims 
that assume vanishing density dependence of the surface tension. 
%is taken for granted.  
However, this assumption applies only for 
the planar interface between the bulk liquid and gas (vacuum) phases where
under mechanical stability, the nucleon density is fixed at the saturation
density and zero, respectively.  A possible constraint on $L$ from empirical 
nuclear masses would thus depend on the adopted density dependence of the 
surface tension.  This situation is similar to the case of the neutron skin 
thickness.

\begin{figure}
\resizebox{0.5\textwidth}{!}{%
  \includegraphics{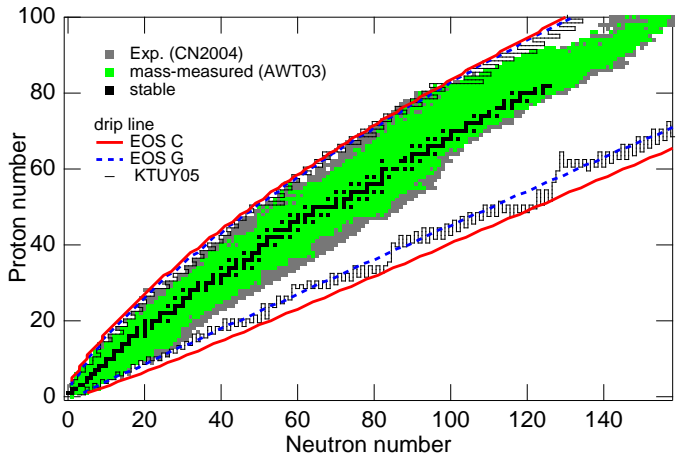}
}
%\vspace{5cm}       % Give the correct figure height in cm
 \caption{(Color online)
The neutron and proton drip lines obtained from the EOS models C and G by
using the macroscopic nuclear model and from a contemporary mass formula
\cite{KTUY}.  The regions filled with squares correspond to empirically
known nuclides \cite{NC2004,AWT}.  From ref.\ \cite{OIK10}.}
\label{chart} 
\end{figure}

    As an important application, we proceed to exhibit the neutron drip line, 
which was calculated from the EOS models C and G within the macroscopic nuclear
model and from the KTUY mass formula \cite{OIK10}.  In fig.\ \ref{chart},
the neutron drip line was 
drawn by identifying nuclides at 
neutron drip with those neighboring to nuclides for which 
$S_{n}=E_B(Z,N)-E_B(Z,N-1)$ and $S_{2n}$ are positive and beyond which at 
least one of them is negative.  As $L$ increases, the calculated drip line 
shifts to the neutron rich side on the chart of nuclides, because for 
larger $L$ we obtain more binding through the surface properties discussed
above.  Consequently, this shift is appreciable for small masses where some 
of the drip nuclei are empirically identified \cite{Notani,Baumann}.  It is an 
interesting 
open question to constrain $L$ from the empirical drip line to be expanded in 
the near future.   Note, however, that the neutron drip line is mainly
determined by a competition between the Coulomb energy and the symmetry energy
coefficient $S_0$, while the present $L$ effect, induced by the surface
properties, is just secondary.

\section{Nuclei in neutron star crusts and nuclear pasta}
\label{pasta}

    Let us turn to a different topic of research, namely, neutron star crusts.
On the basis of ref.\ \cite{OI07}, we will show that the presence of nuclear 
pasta in neutron stars is sensitive to the density symmetry coefficient $L$.

    Nuclear pasta represents exotic shapes of nuclei, which may occur in the
deepest region of the crust \cite{PR,CH}.  In this region, nuclei are 
considered to be 
closely packed in a bcc Coulomb lattice.  Then, the total surface area becomes 
very large.  In order to lower the system energy, it is convenient that the 
spherical nuclei are elongated and fuse into a nuclear rod.  In the presence 
of Coulomb energy, the nuclei cannot have arbitrary shape.  With the density 
increased further, possible changes in nuclear shape are considered to be
rods, slabs, tubes, 
bubbles, and uniform.  In terms of liquid crystals, the rod and tube phases 
are columnar, while the slab phase is smectic A.  Also these pasta phases can 
be regarded as liquid-gas mixed phases.  As we will see, the symmetry energy 
at subnuclear densities controls the crust-core boundary and the presence
of nuclear pasta.

    In describing zero-temperature matter in neutron star crusts, we again use 
the macroscopic nuclear model.  This time, not for a nucleus in vacuum, but 
for a nucleus or bubble in a Wigner-Seitz cell.  New additions are dripped 
neutrons, a neutralizing uniform background of electrons, and the lattice 
energy. 

    For each unit cell, we write the total energy as
\begin{equation}
  W=W_N+W_e+W_C,
   \label{w}
\end{equation}
where $W_N$, $W_e$, and $W_C$ are the nuclear energy, the electron energy,
and the Coulomb energy.  

     As in eq.\ (\ref{e}), the nuclear energy is again 
expressed in the density functional form:
\begin{eqnarray}
  W_N&=&\int_{\rm cell} d^3 r 
    \left\{n({\bf r})w\left[n_n({\bf r}),n_p({\bf r})\right] \right.
   \nonumber \\ &&
   \left.  +m_n c^2 n_n({\bf r})+m_p c^2 n_p({\bf r})
     +F_0 |\nabla n({\bf r})|^2\right\}.
   \label{wn}
\end{eqnarray}
For a spherical nucleus in vacuum, this expression reduces to $E-E_C$
[see eq.\ (\ref{e})].

     The electron energy can be approximated as the energy of an ideal
uniform Fermi gas,
\begin{equation}
    \frac{W_e}{a^3}=\frac{m_e^4c^5}{8\pi^2\hbar^3}
        \{x_e(2x_e^2+1)(x_e^2+1)^{1/2}-\ln[x_e+(x_e^2+1)^{1/2}]\}
    \label{we}
\end{equation}
with
\begin{equation}
    x_e=\frac{\hbar (3\pi^2 n_e)^{1/3}}{m_e c},
    \label{xe}
\end{equation}
where $m_e$ is the electron mass, and $n_e$ is the electron number density
that satisfies the charge neutrality condition,
\begin{equation}
    a^3 n_e= \int_{\rm cell} d^3 r n_p({\bf r}).
    \label{cn}
\end{equation}
We remark that $n_e$ is so high that we can safely ignore inhomogeneity of 
the electron density induced by the electron screening of nuclei or bubbles 
%\cite{WI} 
and the Hartree-Fock corrections to the electron energy.

    The Coulomb energy, which is composed of the proton self-Coulomb energy 
and the lattice energy, can be written as
\begin{equation}
  W_C=\frac12 \int_{\rm cell} d^3 r e[n_p({\bf r})-n_e]\phi({\bf r})
      +\Delta W_1,
    \label{wc}
\end{equation}
where $\phi({\bf r})$ is the electrostatic potential in a Wigner-Seitz
cell, and $\Delta W_1$ is the difference of the rigorous calculation 
\cite{OHY} for a cell in the bcc (triangular) lattice of spherical 
(cylindrical) nuclei or bubbles having sharp surfaces from the 
Wigner-Seitz value, as parametrized in ref.\ \cite{O}.  We take 
into account $\Delta W_1$, which is a less than 1 \% correction,
because $\Delta W_1$ depends sensitively on the dimensionality of the 
lattice. 
%(Note that $\Delta W_1=0$ for the layered lattice of slab nuclei.)

     For nucleon distributions in the Wigner-Seitz cell, we simply 
generalize the parametrization (\ref{ni}) for a nucleus in vacuum into
\begin{equation}
  n_i(r)=\left\{ \begin{array}{lll}
  (n_i^{\rm in}-n_i^{\rm out})
  \left[1-\left(\displaystyle{\frac{r}{R_i}}\right)^{t_i}\right]^3
          +n_i^{\rm out},
         & \mbox{$r<R_i,$} \\
             \\
         n_i^{\rm out},
         & \mbox{$R_i\leq r.$}
 \end{array} \right.
\label{nig} 
\end{equation}
Here $r$ is the distance from the central point, axis, or plane of
the unit cell.  In the case of nuclei, $n_p^{\rm out}=0$, while
in the case of bubbles, $n_p^{\rm in}=0$.

     We finally determine the equilibrium configuration of the system
at given baryon density,
\begin{equation}
  n_b=a^{-3}\int_{\rm cell} d^3 r n({\bf r}).
 \label{nb}
\end{equation}
First, for each of the five inhomogeneous phases, we minimize 
the total energy density $W/a^3$ with respect to the eight
parameters $a$, $n_n^{\rm in}$, $n_n^{\rm out}$, $n_p^{\rm in}$ (for nuclei)
or $n_p^{\rm out}$ (for bubbles), $R_n$, $R_p$, $t_n$, and $t_p$.
This minimization implicitly allows for the stability of the nuclear matter 
region (the region containing protons) with respect to change in the size, 
neutron drip, $\beta$-decay, and pressurization.  In addition to
the five inhomogeneous
phases, we consider a uniform phase of $\beta$-equilibrated, neutral nuclear 
matter.  The energy density of this phase is the sum of the nucleon part 
$nw+m_n c^2 n_n+m_p c^2 n_p$ [see eq.\ (\ref{eos1})] and the electron part 
(\ref{we}).  By comparing the resultant six energy densities, we can 
determine the equilibrium phase of energy density $\rho$.

\begin{figure}
\resizebox{0.5\textwidth}{!}{%
  \includegraphics{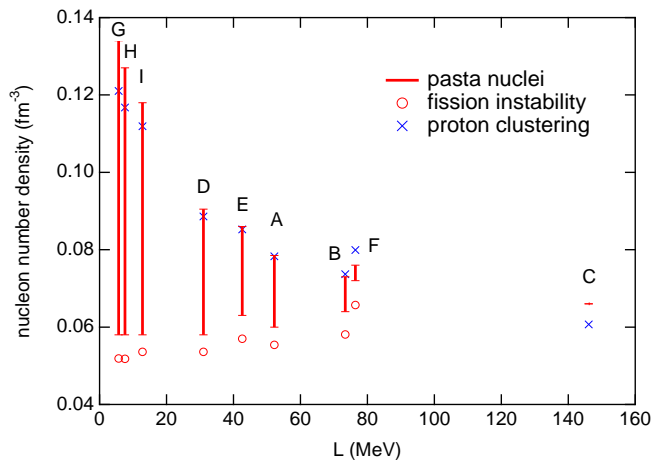}
}
%\vspace{5cm}       % Give the correct figure height in cm
 \caption{(Color online)
The density region containing bubbles and nonspherical
nuclei as a function of $L$, calculated for the EOS models A--I in 
fig.\ \ref{eos}.  For
comparison, the density corresponding to $u=1/8$ in the phase with spherical 
nuclei and the onset density
%, $n(Q)$, 
of proton clustering in uniform nuclear matter
%, which will be discussed in sect.\ \ref{sec:clustering}, 
are also plotted by circles and crosses, respectively.
From ref.\ \cite{OI07}.}
\label{pasta} 
\end{figure}

    In fig.\ \ref{pasta}, we show the resultant density region where pasta 
nuclei are predicted to appear.  We find that the larger $L$, the narrower 
pasta region.  This tendency suggests that for smaller symmetry energy at 
subnuclear densities, protons become more difficult to cluster in uniform 
matter.  

%Judging from the present model calculations, the critical value of 
%$L$ for the presence of pasta nuclei in neutron stars is about 100 MeV.  

%Proton clustering and fission instability

\begin{figure}
\resizebox{0.5\textwidth}{!}{%
  \includegraphics{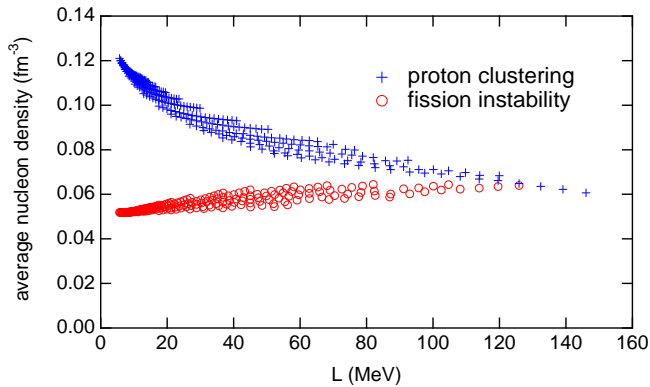}
}
%\vspace{5cm}       % Give the correct figure height in cm
 \caption{(Color online)
The onset density of proton clustering in uniform
nuclear matter calculated from the present EOS models.
For comparison, we plot the density 
corresponding to $u=1/8$ in the phase with spherical nuclei.
%which is taken from fig.\ 6(a).
From ref.\ \cite{OI07}.}
\label{nq} 
\end{figure}

     The lower end of the pasta region can be naively understood from 
fissionlike instability of spherical nuclei.  In a liquid-drop model, it is 
predicted that nuclei tend toward quadrupolar deformations when the Coulomb 
energy is twice as large as the surface energy.  In neutron stars, due to the 
lattice energy, this condition can be essentially met even in equilibrium when 
the volume fraction of nuclei reaches 1/8.  At this volume fraction, the 
baryon density is of order 0.06 fm$^{-3}$ and almost independent of the 
EOS models, as shown in fig.\ \ref{nq}. 

     On the other hand, the upper end of the pasta region can be naively 
understood from proton clustering instability of uniform matter.  The tendency 
to proton clustering can be measured by the sign change of the effective 
potential between proton density fluctuations.  The driving force of proton 
clustering is the symmetry energy at subnuclear densities.  In fact, for 
larger $L$, the system keeps homogeneous down to lower density, as shown in 
fig.\ \ref{nq}.

     Comparing the onset densities of proton clustering and fission 
instability with the equilibrium calculations of the pasta region,  we find 
that the onset densities are a good measure of the pasta region (see fig.\ 
\ref{pasta}).  Judging 
from fig.\ \ref{nq}, the critical value of $L$ for the presence of pasta 
nuclei in neutron stars is about 100 MeV.

%     We remark the claim of Chen et al.\ that a parabolic approximation for 
%the {\em bulk energy} with respect to neutron excess does not duly predict 
%an $L$ dependence of the crust-core boundary.  In the present EOS models, we 
%assume the parabolic form solely for the potential energy part as in 
%eq.\ (\ref{eos1}), but the kinetic energy part does include the terms beyond 
%the parabolic form.  The assumption adopted here is partially justified by 
%variational calculations of Lagaris and Pandparihande \cite{LP}.

      Recently, more complicated structures with intersecting rods have been 
predicted by calculations beyond the Wigner-Seitz approximation 
\cite{QMD,SHF1,SHF2,Sebille1,Sebille2,Bastian}.  
On the other hand, a liquid-drop approach \cite{NOY,NIO} can be used for the 
purpose 
of examining the possible occurrence of a periodic bicontinuous structure,
namely, gyroid, which is known to occur in polymer systems \cite{bafre99}.  
Energetically, 
this structure is to be reckoned with seriously judging from the evaluated
energy difference from the ground state at subnuclear densities, but even 
with shape-dependent curvature effects included, no density region 
where the gyroid is most stable was found.

     At finite temperatures, pasta nuclei do not always have perfect structure.
%as demonstrated by quantum molecular dynamics and time-dependent
%Hartree-Fock calculations [].  
In fact, they thermally fluctuate just like
molecular liquid crystals.  At typical temperatures of neutron star 
interiors, however, the amplitude of the displacements involved is smaller 
than internuclear spacing \cite{WIS}.

\begin{figure}
\resizebox{0.5\textwidth}{!}{%
  \includegraphics{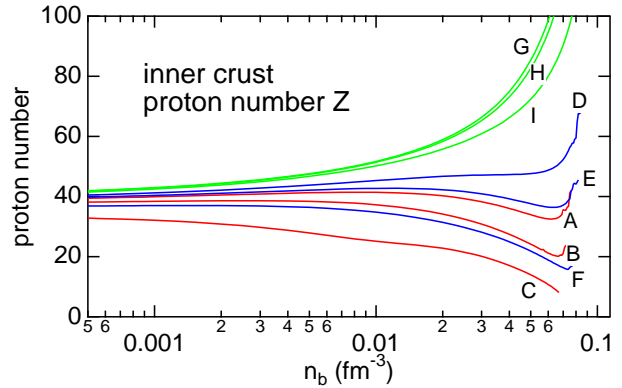}
}
%\vspace{5cm}       % Give the correct figure height in cm
 \caption{(Color online)
The charge number of spherical nuclei as a function of
$n_b$, calculated for the EOS models A--I.
From ref.\ \cite{OI07}.}
\label{Z} 
\end{figure}

    We conclude this section by showing the size of spherical nuclei in the 
inner crust estimated for various EOS models within the macroscopic nuclear 
model.  We find from fig.\ \ref{Z} that the larger $L$, the smaller size.  
This tendency 
suggests that for larger $L$ or, equivalently, smaller symmetry energy at 
subnuclear density, the density of dripped neutrons becomes larger and hence 
the surface tension becomes smaller.  This is essential to possible 
constraints on $L$ from neutron star asteroseismology as will be discussed 
in the next section.

\section{Possible constraints on $L$}
\label{L}

%briefly describe possible constraints from exps on unstable nuclei

     In this section we consider possible constraints on $L$ from empirical 
data via the macroscopic nuclear model.  As discussed in sect.\ 5, experimental
data on masses of unstable nuclei could give a stringent constraint on $L$
if the model were free from systematic errors associated with the description 
of the isospin-dependent surface properties.  In fact, constraints on $L$
that were obtained from the mass data in earlier publications scatter 
\cite{Newton}.
All we can conclude at this stage is that a very small $L$ cannot reproduce
the empirical isotope dependence of the two-proton separation energy $S_{2p}$
depicted in fig.\ \ref{S2p}.  On the other hand, future systematic data 
associated with the size of unstable nuclei are expected to help constrain 
$L$ as discussed in Secs.\ 3 and 4.

    We finally turn to QPOs in giant flares from SGRs and their possible
relation with crustal torsional oscillations.  Usually, SGRs are considered
to be magnetars, i.e., neutron stars with surface magnetic fields of order
$10^{15}$ G.  About a decade ago, one of them exhibited a giant flare and 
fortunately, its X-ray afterglow was detected by the Rossi X-ray Timing
Explorer \cite{WS06}.  It turns out that the afterglow oscillates 
quasi-periodically.

   Steiner and Watts \cite{SW09} tried to explain these QPOs in terms of 
crustal shear 
modes.  They succeeded in reproducing some of the measured frequencies,
but the analysis is model dependent.  This is because the shear modulus is 
controlled by the charge number of neutron-rich nuclei that constitute the 
crust, and the charge number is predicted to be dependent on $L$ as shown
in fig.\ \ref{Z}.

   With this $L$ dependence of the charge number in mind, we evaluated the
frequency of the fundamental mode of crustal torsional oscillation by ignoring 
and allowing for the effects of neutron superfluidity \cite{SNIO12,SNIO13}.  
For such evaluations, we first consider the equilibrium neutron 
star configurations.  Since the magnetic energy is much smaller than the 
gravitational binding energy even for magnetars, we can neglect the 
deformation due to the magnetic pressure. Additionally, since the magnetars 
are relatively slowly rotating, we can also neglect the rotational effect.  
Hereafter, therefore, we consider spherically symmetric neutron stars, whose 
structure is described by the solutions of the well-known 
Tolman-Oppenheimer-Volkoff (TOV) equations.  In this case, the metric 
can be expressed in terms of the spherical polar coordinates $r$, 
$\theta$, and $\phi$ as
\begin{equation}
 ds^2 = -{\rm e}^{2\Phi}dt^2 + {\rm e}^{2\Lambda}dr^2 + r^2 d\theta^2 
+ r^2\sin^2\theta\, d\phi^2,
\label{metric}
\end{equation}
where $\Phi$ and $\Lambda$ are functions of $r$.  (Hereafter, we use units in
which $G=c=1$.)  We remark that $\Lambda(r)$ 
is associated with the mass function 
\begin{equation}
m(r)=\int_0^r dr' 4\pi r'^2\rho (r') 
\end{equation}
as ${\rm e}^{2\Lambda}=[1-2m(r)/r]^{-1}$.

     To solve the TOV equations, one generally uses the zero-temperature 
EOS, i.e., the pressure $p$ as a function of the mass density $\rho$.  
For matter in the crust, we use the same EOS models as described above.  
Unfortunately, the core EOS is still uncertain in the absence of clear 
understanding of the constituents and their interactions both in vacuum and in 
medium.  Since we will focus on shear torsional oscillations that occur in 
the crust, we can effectively describe such uncertainties in the core EOS by 
solely setting the star's mass $M$ and radius $R$ as free parameters, without 
using specific models for the core EOS.  In fact, for various sets of $M$ 
and $R$, we systematically construct the equilibrium configuration of the 
crust by integrating the TOV equations with the crust EOS from the star's 
surface all the way down to the crust-core boundary. 
%as in ref.\ [IS1997].
This is a contrast to the usual way of constructing a star by initially 
giving a value of the central mass density and then integrating the TOV 
equations with a specific model for the EOS from the star's center to surface.
Hereafter, we will consider $1.4\le M/M_\odot\le 1.8$ and 10 km $\le R\le$ 14 
km as typical values of $M$ and $R$.  Such choice of $M$ and $R$ can duly 
encapsulate uncertainties of the core EOS.

    Generally, a restoring force for shear torsional oscillations is 
provided by shear stress,  which comes from the elasticity of the
oscillating body and is characterized by the shear modulus $\mu$.  
In the case of torsional oscillations in the crust of a neutron star, the 
shear modulus is determined by the lattice energy of the Coulomb crystal
that constitutes the crust.  Since the crystal is generally considered to be 
of bcc type 
%(an fcc lattice might occur in place of bcc in the crust as suggested by 
%recent Thomas-Fermi calculations \cite{Okamoto2012}), 
one can use the corresponding shear modulus, which is calculated for $Ze$ 
point charges of number density $n_i$ in the uniform neutralizing background
as 
\begin{equation}
\mu = \frac{0.1194n_i (Ze)^2}{a}, \label{eq:shear}
\end{equation}
where $a=(3Ze/4\pi n_e)^{1/3}$ is the radius of a Wigner-Seitz cell 
\cite{SHOII1991}.  Note that this formula is derived 
in the limit of zero temperature from Monte Carlo calculations for 
the shear modulus averaged over all directions \cite{OI1990}. 
%As shown in \cite{SNIO12}, 

     The shear modulus depends strongly on 
the value of $L$, which comes mainly from the $L$ dependence of the 
calculated $Z$ \cite{OI07}.  It is natural that one should take 
into account the shear modulus in pasta phases, if present, but hereafter 
we simply assume $\mu=0$ for the pasta phases.
%, as in \cite{GNHL2011,SNIO2012,SNIO2013}.  
This is because the shear modulus in the pasta phases except a phase of 
spherical bubbles has at least one direction in which the system is invariant
with respect to translation and hence is expected to be significantly smaller 
than that in a phase of spherical nuclei \cite{PP98}.  Under this 
assumption, we have only to consider the shear torsional oscillations that 
are excited within a crustal region of spherical nuclei.
%, i.e., for $n_b\le n_1$. 
Anyway, the constraint on $L$ that will be given below can be considered 
to be robust, because the pasta region as shown in fig.\ \ref{pasta} is 
highly limited given the resulting constraint on $L$.

     We now consider the shear torsional oscillations on the equilibrium 
configuration of the crust of a neutron star described above. In order to
determine the frequencies, we adopt the relativistic Cowling approximation, 
i.e., we neglect the metric perturbations on eq.\ (\ref{metric}) by 
setting $\delta g_{\mu\nu}=0$.  In fact, one can consider the shear torsional 
oscillations with satisfactory accuracy even with the relativistic Cowling 
approximation, because the shear torsional oscillations on a spherically 
symmetric star are incompressible and thus independent of the density 
variation during such oscillations.  Additionally, due to the spherically 
symmetric nature of the background, we have only to consider the 
axisymmetric oscillations.  Then, the only non-zero perturbed matter 
quantity is the $\phi$ component of the perturbed four-velocity, 
$\delta u^{\phi}$, which can be written as
\begin{equation}
   \delta u^{\phi} = {\rm e}^{-\Phi}\partial_t {\cal Y}(t,r)\frac{1}{\sin\theta}\partial_\theta P_{\ell}(\cos\theta), 
\end{equation}
where $\partial_t$ and $\partial_\theta$ denote the partial derivatives with 
respect to $t$ and $\theta$, respectively, while $P_\ell(\cos\theta)$ is 
the $\ell$-th order Legendre polynomial.  We remark that ${\cal Y}(t,r)$ 
characterizes the radial dependence of the angular displacement of a
matter element. By assuming that the perturbation variable ${\cal Y}(t,r)$ has
such a harmonic time dependence as ${\cal Y}(t,r)={\rm e}^{{\rm i}\omega 
t}{\cal Y}(r)$, the perturbation equation that governs the shear 
torsional oscillations can be derived from the linearized equation of 
motion as \cite{ST83}
\begin{eqnarray}
 {\cal Y}'' + \left[\left(\frac{4}{r}+\Phi'-\Lambda'\right)+\frac{\mu'}{\mu}\right]{\cal Y}'  
\nonumber \\
      + \left[\frac{H}{\mu}\omega^2{\rm e}^{-2\Phi}-\frac{(\ell+2)(\ell-1)}{r^2}\right]{\rm e}^{2\Lambda}{\cal Y} &=& 0,
 \label{eq:perturbation}
\end{eqnarray}
where $H$ is the enthalpy density defined as $H\equiv \rho+p$ with the 
mass density $\rho$ and pressure $p$ as described in the previous section,
and the prime denotes the derivative with respect to $r$.  Note that under 
the present definition of $H$, effects of neutron superfluidity are
essentially ignored as in ref.\ \cite{SNIO12}, while such effects will be 
included below as in ref.\ \cite{SNIO13}.

     Once appropriate boundary conditions are imposed, the problem to 
solve reduces to an eigenvalue problem with respect to $\omega$.  Since 
there is no matter outside the star, we adopt the zero-torque condition at the 
star's surface.  Meanwhile, since there is no traction force in the 
region with $\mu=0$, we adopt the zero-traction condition at the position 
where spherical nuclei disappear in the deepest region of the crust.  In 
practice, one can show that both conditions reduce to ${\cal Y}'=0$.

\begin{figure}
\resizebox{0.5\textwidth}{!}{%
  \includegraphics{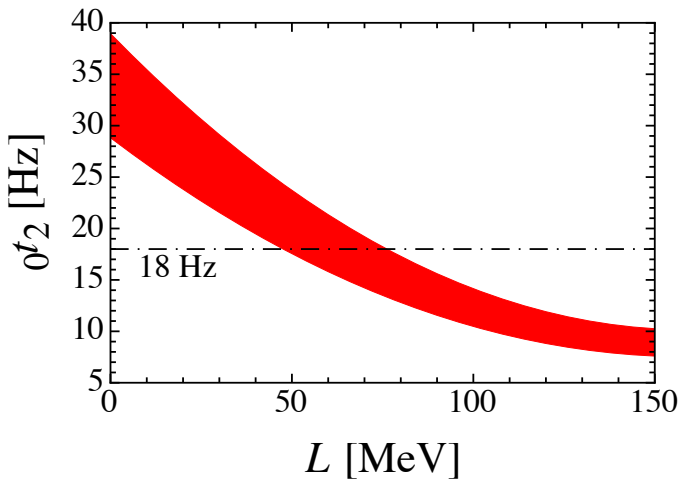}
}
%\vspace{5cm}       % Give the correct figure height in cm
 \caption{(Color online)
$_0 t_2$ as a function of $L$ for $10$ km $\le R \le14$ km and 
$1.4M_\odot\le M\le 1.8M_\odot$.
The horizontal dot-dashed line denotes the lowest QPO 
frequency observed from SGR 1806-20 \cite{WS06}.
From ref.\ \cite{SNIO12}.}
\label{allowed} 
\end{figure}

     For the EOS models A--I, we then calculate $_0 t_2$, namely, the 
frequency of the mode with zero radial node and spherical harmonics $\ell=2$,
which is theoretically the lowest among various eigenfrequencies of the
torsional oscillations.  By interpolating the results for $_0 t_2$ and 
assuming $1.4\le M/M_\odot\le 1.8$ and 10 km $\le R\le$ 14 km,
we obtain an allowed region as in fig.\ \ref{allowed}.
We can clearly see the $L$ dependence of the frequency, while the width 
comes from various sets of the star's radius and mass.  These features occur
because $_0 t_2$ is basically determined by the ratio of the shear velocity 
$\sqrt{\mu/H}$ over the oscillation path length through the crust.  
Note that the lower (upper) boundary of the allowed region in fig.\ 
\ref{allowed} corresponds to the largest (smallest) and heaviest (lightest)
case.  In these 
calculations, we ignore the effects of neutron superfluidity and pasta 
elasticity, which act to enhance the frequency.  As a result, these 
calculations are expected to provide a lower limit of $L$, hereafter referred 
to as $L_{\rm min}$, under the assumption that the observed QPOs in SGR
giant flares arise from the torsional oscillations in neutron star crusts. 
In fact, since, under this assumption, $_0 t_2$ would be equal to or even 
lower than the lowest QPO frequency observed, we obtain $L_{\rm min}$ of 
order 50 MeV, which is fairly stringent given that $L$ is still uncertain 
\cite{Newton}.  Note that we ignore electron screening and finite-size 
effects on charge \cite{BBP,HH}, which act to reduce the shear modulus
and hence $_0 t_2$, as well as nuclear shell and pairing effects \cite{DSB}, 
which act to shift the nuclear charge number by keeping the local average
almost unchanged.  Such ignorance could modify $L_{\rm min}$.

%SUPERFLUIDITY

     Now, we take into account the effect of neutron superfluidity on the 
shear torsional oscillations \cite{SNIO13}.  
In general, it is considered that neutrons 
confined in the nuclei start to drip therefrom when the mass density becomes 
higher than $\sim 4\times 10^{11}$ g cm$^{-3}$.  Then, some of the dripped 
neutrons can behave as a superfluid.  
%Although the behavior of the dripped neutrons is not fully understood, 
A significant fraction of the dripped neutrons may move non-dissipatively 
with protons in the nuclei as a result of Bragg scattering off the bcc 
lattice of the nuclei; only neutrons in the conduction band can freely
flow with respect to the lattice.  In fact, the recent band calculations 
beyond the Wigner-Seitz approximation by Chamel \cite{Chamel12} show that 
the superfluid density, which is 
defined here as the density of neutrons unlocked to the motion of protons in 
the nuclei, depends sensitively on the baryon density above neutron drip
and that a considerable portion of the dripped neutrons can be 
locked to the motion of protons in the nuclei.  On the other hand, since the 
shear torsional oscillations are transverse, the remaining superfluid 
neutrons, whose low-lying excitations are longitudinal, do not contribute
to such oscillations \cite{PCR10}.

     We build the effect of neutron superfluidity into the 
effective enthalpy density $\tilde{H}$, which can be determined by subtracting 
the superfluid mass density from the total enthalpy density $H$ in eq.\ 
(\ref{eq:perturbation}) that fully contains the contributions of 
the superfluid neutrons as well as the nuclei and companions. 
%\cite{IB-II}.  
Since we assume that the temperature of neutron star matter is zero, the 
baryon chemical potential $\mu_b$ can be expressed as 
$\mu_b=H/n_b$.  Thus, one can write down 
\begin{equation}
   \tilde{H} = \left(1-\frac{N_s}{A}\right)H,
\end{equation}
where $N_s$ denotes the number of neutrons in a Wigner-Seitz cell that do not 
comove with protons in the nucleus, while $A$ is the total nucleon number 
in the Wigner-Seitz cell.  Finally, substituting $\tilde{H}$ for $H$ in 
eq.\ (\ref{eq:perturbation}), one can determine the frequencies of 
the shear torsional oscillations, which include the effect of 
neutron superfluidity in a manner that depends on the value of $N_s$. 
Hereafter, we will assume that $N_s$ comes entirely from the dripped neutron 
gas.  Even so, it is still uncertain how much fraction of the
dripped neutrons behave as a superfluid. Thus, 
%as in \cite{SNIO13}, 
we introduce a new parameter $N_s/N_d$, where $N_d$ is the number of the 
dripped neutrons in the Wigner-Seitz cell.  For $N_s/N_d=0$, 
all the dripped neutrons behave as normal matter and contribute to 
the shear motion, while for $N_s/N_d=1$, all the dripped neutrons 
behave as a superfluid.  We remark that $N_d-N_s$ denotes the 
number of the dripped neutrons bound to the nucleus.  Typically, the value of 
$N_s/N_d$ depends on the density inside a neutron star \cite{Chamel12}, 
but the case of $N_s/N_d=0$ in the whole crust is closer to the typical 
behavior than the case of $N_s/N_d=1$.

\begin{figure}
\resizebox{0.5\textwidth}{!}{%
  \includegraphics{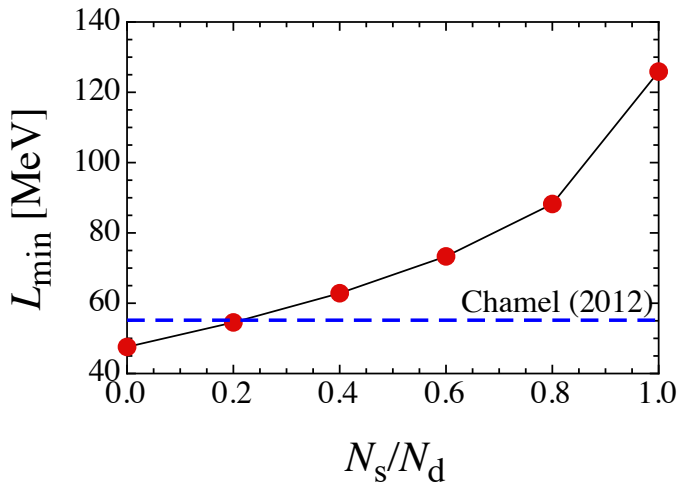}
}
%\vspace{5cm}       % Give the correct figure height in cm
 \caption{(Color online)
$L_{\rm min}$ as a function of $N_s/N_d$, the fraction of 
superfluid neutrons.  The horizontal broken line corresponds to the 
result from the band calculations of $N_s/N_d$ by Chamel 
\cite{Chamel12}.
From ref.\ \cite{SNIO13}.}
\label{fig:Lmin} 
\end{figure}

     We proceed to show how neutron superfluidity affects $L_{\rm min}$.  
For constant values of $N_s/N_d$, we calculate $_0 t_2$ by following the same 
line of argument as in the absence of neutron superfluidity 
(i.e., $N_s/N_d=0$) and therefrom obtain $L_{\rm min}$, which is illustrated 
in fig.\ \ref{fig:Lmin}.  One can observe that the value of $L_{\rm min}$, 
which is 47.6 MeV in the case of $N_s/N_d=0$, can be as 
large as 125.9 MeV in its presence (i.e., $0<N_s/N_d\leq1$).
In addition, we exhibit $L_{\rm min}=55.2$ MeV, the result from 
the realistic band calculations of $N_s/N_d$ in ref.\ \cite{Chamel12}. 
%the broken line in fig.\ \ref{fig:0t2Lrate}.  
This $L_{\rm min}$ is expected to give a reliable constraint on $L$, 
although a possible $L$ dependence of $N_s/N_d$ through the band structure
remains to be investigated.

\begin{figure}
\resizebox{0.5\textwidth}{!}{%
  \includegraphics{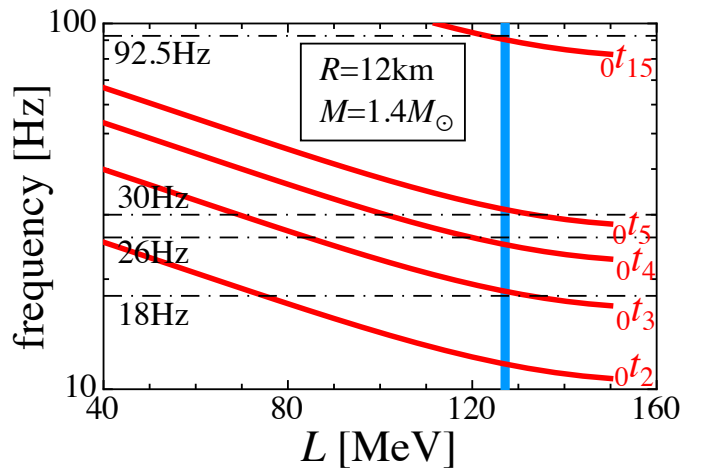}
}
%\vspace{5cm}       % Give the correct figure height in cm
 \caption{(Color online)
Comparison of the calculated frequencies of torsional 
oscillations (solid lines) with the QPO frequencies observed in SGR 1806-20 
(dot-dashed lines), where the stellar model adopted in the calculations
is $M/M_\odot=1.4$ and $R=12$ km.  The vertical line corresponds to
the value of $L$ that is consistent with the observations.
From ref.\ \cite{SNIO13}.}
\label{fig:fit-M14R12} 
\end{figure}

    Instead of just considering $L_{\rm min}$, we can obtain a more
stringent constraint on $L$ by fitting the predicted frequencies of 
fundamental torsional oscillations with different values of $\ell$ to the 
low-lying QPO frequencies observed in SGRs.  To this end, we use the 
values of $N_s/N_d$ derived by Chamel \cite{Chamel12}.  In the present 
analysis, we focus on the observed QPO frequencies lower than 100 
Hz, i.e., 18, 26, 30, and 92.5 Hz in SGR 1806-20 and 28, 54, and 84 Hz in SGR 
1900+14 \cite{WS06}.  In fact, the even higher observed frequencies would 
be easier to explain in terms of multipolar fundamental and overtone 
frequencies.  Because of the small interval between the observed frequencies 
26 and 30 Hz in SGR 1806-20, it is more difficult to explain the QPO 
frequencies observed in SGR 1806-20 than those in SGR 1900+14. 
%\cite{Sotani2007}.  
If one identifies the lowest frequency in SGR 1806-20 
(18 Hz) as the fundamental torsional oscillation with $\ell=3$, 
%as in ref.\ \cite{Sotani2011b}, 
one can reasonably explain 26, 30, and 92.5 Hz in terms of those with 
$\ell=4$, 5, and 15.  In the case of the typical neutron star model with 
$M=1.4M_\odot$ and $R=12$ km, we compare the predicted frequencies with 
the observed ones as shown in fig.\ \ref{fig:fit-M14R12}.  One can observe 
from this figure that the best value of $L$ to reproduce the observed 
frequencies is $L=127.1$ MeV, where the calculated frequencies, 18.5 Hz 
($\ell=3$), 24.9 Hz ($\ell=4$), 31.0 Hz ($\ell=5$), and 90.3 Hz 
($\ell=15$), are within less than $5\%$ deviations from the observations.

    Let us now extend the analysis to different stellar models and to SGR
1900+14.  We find that the QPOs observed in SGR 1806-20 can be 
explained in terms of the eigenfrequencies of the same $\ell$ within 
similar deviations even for different stellar models except for the case 
with $M=1.4M_\odot$ and $R=10$ km, while the low-lying QPOs observed in 
SGR 1900+14 can be similarly explained in terms of the fundamental 
torsional oscillations with $\ell=4$, 8, and 13.  As a result, the 
allowed region of $L$ where the QPO frequencies observed in SGR 1806-20 
and in SGR 1900+14 are reproducible simultaneously lies in the range 
101.0 MeV$\le L\le 130.1$ MeV, as long as the oscillating neutron stars 
have mass and radius ranging $1.4\le M/M_\odot\le 1.8$ and 10 km 
$\le R\le$ 14 km.  It is interesting to compare this constraint with 
various experimental constraints on $L$ \cite{Newton}, which have yet to 
converge but seemingly favor smaller $L$.  Note that if we extend
the mass range to, e.g., $1.0\le M/M_\odot\le 1.8$, this constraint
would be broader with the lower bound unchanged, while the values of 
$L_{\rm min}$ in fig.\ \ref{fig:Lmin} would be unchanged.  This is
because of larger $_0 t_2$ for smaller $M$.

    In principle, there are many other ways of identifying the observed
QPOs.  For example, magnetic oscillations and magneto-elastic 
oscillations \cite{PL13,Gabler} have been already invoked as such candidates.  
Our analyses, 
however, imply that as good a reproduction of the observed low-lying 
frequencies as shown in fig.\ \ref{fig:fit-M14R12} would be desired. 
It is an open issue to ask if there could be a new way of reasonable
identification that predicts a lower $L$.

\section{Epilogue}
\label{conc}

    As we have shown by using the macroscopic nuclear model, the parameter 
$L$ controlling the density dependence of the symmetry energy is closely 
related to the size and mass of unstable nuclei in laboratories and to the 
pasta region and shear modes in neutron star crusts.  Ongoing and future 
developments of neutron star observatories and RI beam facilities are thus 
expected to help determining $L$ sufficiently well.  For such determination, 
systematic errors involved in connecting $L$ with observables would have 
to be duly taken into account, no matter whether the macroscopic nuclear 
model is used or not.

%Isospin dependent phenomena such as dipole resonances and 
%isoscaling/isospin diffusion are more or less sensitive to the parameter L.  
%Thank you.

\section*{Acknowledgments}

We are indebted to our collaborators of the works underlying the present 
contribution, namely, B. Abu-Ibrahim, A. Kohama, H. Koura, K. Nakazato, 
and H. Sotani.  This work was supported in part by Grants-in-Aid for 
Scientific Research on Innovative Areas through No.\ 24105008 provided by 
MEXT.

% BibTeX users please use
% \bibliographystyle{}
% \bibliography{}

\end{document}